\newcommand*{\ATLASLATEXPATH}{latex/}
\documentclass[cernpreprint,texlive=2011,txfonts,UKenglish]{\ATLASLATEXPATH atlasdoc}
\pdfoutput=1
\usepackage{\ATLASLATEXPATH atlaspackage}
\usepackage{\ATLASLATEXPATH atlasbiblatex}
\usepackage{\ATLASLATEXPATH atlascontribute}
\usepackage{\ATLASLATEXPATH atlasphysics}
\graphicspath{{figures/}}
\usepackage{multirow}
\usepackage{bigdelim}
\usepackage{paper-defs}
\AtlasTitle{Measurement of the ${\boldsymbol{t\bar{t}Z}}$ and
${\boldsymbol{t\bar{t}W}}$ production cross sections\\ in multilepton final
states using 3.2 fb$^{-1}$ of $\boldsymbol{pp}$ collisions\\ at $\sqrt{s}$ = 13
Te\kern -0.1em V with the ATLAS detector}

\PreprintIdNumber{CERN-EP-2016-185}
\AtlasDate{\today}
\arXivId{1609.01599}
\AtlasDOI{10.1140/epjc/s10052-016-4574-y}
\AtlasJournalRef{\EPJ\ C77 (2017) 40}

\AtlasAbstract{A measurement of the $t\bar{t}Z$ and $t\bar{t}W$ production
cross sections in final states with either two same-charge muons, or three or four
leptons (electrons or muons) is presented.  The analysis uses a data sample of
proton--proton collisions at $\sqrt{s} = 13$ Te\kern -0.1em V recorded with the
ATLAS detector at the Large Hadron Collider in 2015, corresponding to a total
integrated luminosity of \mbox{3.2 fb$^{-1}$}.  The inclusive cross sections
are extracted using likelihood fits to signal and control regions, resulting in
$\sigma_{t\bar{t}Z} = 0.9 \pm 0.3$ pb and $\sigma_{t\bar{t}W} = 1.5 \pm 0.8$ pb,
in agreement with the Standard Model predictions.}

\hypersetup{pdftitle={ATLAS document},pdfauthor={The ATLAS Collaboration}}

\begin{document}
\maketitle
\section{Introduction}
\label{s:intro}

At the Large Hadron Collider (LHC), top quarks are copiously produced in
quark--antiquark pairs (\ttbar). This process has been extensively studied in proton--proton collisions at 
$7$ and $8\,\TeV$, and recently at $13\,\TeV$~\cite{TOPQ-2015-09, CMS-TOP-15-003}
centre-of-mass energy.  Measurements of the associated production of \ttbar
with a $Z$ boson (\ttZ) allow the extraction of information about the neutral-current coupling of the top quark.  The production rate of a top-quark pair with a
massive vector boson could be altered in the presence of physics beyond the
Standard Model (SM), such as vector-like quarks~\cite{AguilarSaavedra:2009es,Aguilar-Saavedra:2013qpa}, strongly coupled Higgs bosons~\cite{Perelstein:2005ka} 
or technicolour~\cite{Chivukula:1992ap, Chivukula:1994mn, Hagiwara:1995jx, Mahanta:1996ng, Mahanta:1996qe},
and therefore the measurements of \sttZ and \sttW are important
checks of the validity of the SM at this new energy regime.  
The \ttZ and \ttW
processes have been established by ATLAS~\cite{TOPQ-2013-05} and
CMS~\cite{CMS-TOP-14-021} using the Run-1 dataset at $\sqrt{s} = 8\,\TeV$, with
measured cross sections compatible with the SM prediction and having uncertainties of
$\sim\!\!30\%$.  At $\sqrt{s} = 13\,\TeV$, the SM cross sections of the \ttZ and
\ttW processes increase by factors of $3.5$ and $2.4$, respectively, compared
to $\sqrt{s} = 8\,\TeV$. The cross sections, computed at next-to-leading-order
(NLO) QCD precision, using \MGMCatNLO (referred to in the
following as \MGAMC), are $\sttZ = 0.84\,\text{pb}$ and $\sttW =
0.60\,\text{pb}$ with an uncertainty of $\sim\!\!12\%$~\cite{Frixione:2015zaa, Alwall:2014hca}, primarily due to higher-order
corrections, estimated by varying the renormalisation and
factorisation scales.

This paper presents measurements of the \ttZ and \ttW cross sections using
\lumi of proton--proton ($pp$) collision data at $\sqrt{s} = 13\,\TeV$
collected by the ATLAS detector in 2015.  The final states of top-quark pairs
produced in association with a $Z$ or a $W$ boson comprise up to four isolated,
prompt leptons.\footnote{In this paper, lepton is used to denote electron or
muon, and prompt lepton is used to denote a lepton produced in a $Z$ or $W$
boson or $\tau$-lepton decay.}  Decay modes with two same-sign (SS) charged
muons, or three or four leptons are considered in this analysis.  The analysis
strategy follows the strategy adopted for the $8\,\TeV$
dataset~\cite{TOPQ-2013-05}, excluding the lower sensitivity SS dilepton
channels.  Table~\ref{tab:intro-channels} lists the analysis channels and the
targeted decay modes of the \ttZ and \ttW processes.  Each channel is divided
into multiple analysis regions in order to enhance the sensitivity to the
signal.  Simultaneous fits are performed to the signal regions and selected
control regions in order to extract the cross sections for \ttZ and \ttW
production.  Additional validation regions are defined to check that the
background estimate agrees with the data and are not used in the fit.

\begin{table}[htbp]
\centering
\caption{\label{tab:intro-channels} List of \ttW and \ttZ decay modes and
analysis channels targeting them.\vspace{1ex}} 
\begin{tabular}{cccc}
\toprule
Process & \ttbar decay & Boson decay & Channel\\
\midrule
\multirow{2}{*}{$\ttW$} 
&  $(\mu^{\pm}\nu b) (q\bar{q} b) $ & $\mu^{\pm}\nu$ & SS dimuon\\
& $ (\ell^{\pm}\nu b) (\ell^{\mp}\nu b)$ & $\ell^{\pm}\nu$ & Trilepton\\
\midrule
\multirow{2}{*}{\ttZ} 
& $(\ell^{\pm}\nu b) (q\bar{q} b)$ & $ \ell^{+}\ell^{-}$ & Trilepton\\
& $(\ell^{\pm}\nu b) (\ell^{\mp} \nu b)$ & $ \ell^{+}\ell^{-}$ & Tetralepton\\
\bottomrule
\end{tabular}
\end{table}

\section{The ATLAS detector}
\label{s:atlas}

The ATLAS detector~\cite{PERF-2007-01} consists of four main subsystems: an
inner tracking system, electromagnetic (EM) and hadronic calorimeters, and a
muon spectrometer (MS).  The inner detector (ID) consists of a high-gra\-nu\-la\-ri\-ty
silicon pixel detector, including the newly installed Insertable
B-Layer~\cite{ATL-INDET-PUB-2015-001}, which is the innermost layer of the
tracking system, and a silicon microstrip tracker, together providing precision
tracking in the pseudorapidity\footnote{ATLAS uses a right-handed coordinate
system with its origin at the nominal interaction point (IP) in the centre of
the detector and the $z$-axis along the beam pipe. The $x$-axis points from the
IP to the centre of the LHC ring, and the $y$-axis points upward. Cylindrical
coordinates $(r,\phi)$ are used in the transverse plane, $\phi$ being the
azimuthal angle around the $z$-axis. The pseudorapidity is defined in terms of
the polar angle $\theta$ as $\eta=-\ln\tan(\theta/2)$.} range $|\eta|<2.5$ and
of a transition radiation tracker covering $|\eta|<2.0$. All the systems are immersed in a
\SI{2}{T} magnetic field provided by a superconducting solenoid.  The EM
sampling calorimeter uses lead and liquid argon (LAr) and is divided into
barrel ($|\eta|<1.475$) and endcap ($1.375<|\eta|<3.2$) regions.  Hadron
calorimetry is provided by a steel/scintillator-tile calorimeter, segmented
into three barrel structures, in the range $|\eta|<1.7$, and by two copper/LAr
hadronic endcap calorimeters that cover the region $1.5<|\eta|<3.2$. The solid
angle coverage is completed with forward copper/LAr and tungsten/LAr
calorimeter modules, optimised for EM and hadronic measurements respectively,
covering the region $3.1<|\eta|<4.9$. The muon spectrometer measures the
deflection of muon tracks in the range $|\eta|<2.7$ using multiple layers of
high-precision tracking chambers located in toroidal magnetic fields.
The field integral of the toroids ranges between $2.0$ and \SI{6.0}{Tm} for
most of the detector.  The muon spectrometer is also instrumented with separate
trigger chambers covering $|\eta|<2.4$.  A two-level trigger system, using
custom hardware followed by a software-based trigger level, is used to reduce the event
rate to an average of around \SI{1}{kHz} for offline storage. 

\section{Data and simulated event samples}
\label{s:samples}

The data were collected with the ATLAS detector during 2015 with a 
bunch spacing of \SI{25}{ns} and a mean number of $14$ $pp$ 
interactions per bunch crossing (pile-up).
With strict data-quality requirements, the integrated luminosity considered
corresponds to \lumi with an uncertainty of $2.1\%$~\cite{Aaboud:2016hhf}. 

Monte Carlo simulation samples (MC) are used to model the expected signal and
background distributions in the different control, validation and signal regions
described below. The
heavy-flavour decays involving $b$- and $c$-quarks, 
particularly important to this measurement, are modelled
using the \EVTGEN~\cite{EvtGen} program, except for processes modelled using the
\SHERPA generator. In all samples the top-quark mass is set to $172.5\,\GeV$ and
the Higgs boson mass is set to $125\,\GeV$. The response of the detector to
stable\footnote{A particle is considered stable if $c\tau \ge 1$ cm.} particles
is emulated by a dedicated simulation~\cite{SOFT-2010-01} based either fully on
\GEANT~\cite{geant} or on a faster parameterisation~\cite{ATL-PHYS-PUB-2010-013}
for the calorimeter response and \GEANT for other detector systems. To account for
additional $pp$ interactions from the same and close-by bunch crossings, a set
of minimum-bias interactions generated using \PYTHIA v8.210~\cite{Sjostrand:2014zea}, 
referred to as \PYTHIA 8 in the following, 
with the \tune{A2}~\cite{ATL-PHYS-PUB-2011-014} set of tuned MC parameters (\tune{A2} tune)
is superimposed on the
hard-scattering events.  In order to reproduce the same pile-up levels present in
the data, the distribution of the number of additional $pp$ interactions in the
MC samples is reweighted to match the one in the data. All samples are processed
through the same reconstruction software as the data. Simulated events are
corrected so that the object identification, reconstruction and trigger
efficiencies, energy scales and energy resolutions match those determined from
data control samples.

The associated production of a top-quark pair with one or two vector bosons is
generated at leading order (LO) with \MGAMC\ interfaced to \PYTHIA 8, with up to
two (\ttW), one (\ttZ) or no ($\bgttWW$) extra partons included in the matrix
elements. The $\gamma^{*}$ contribution and the \Zgamstar interference are 
included in the \ttZ samples. 
The \tune{A14} ~\cite{ATL-PHYS-PUB-2014-021} set of tuned MC parameters (\tune{A14} tune) is used together with the 
\pdf{NNPDF2\!.\!3LO} parton distribution function (PDF) set~\cite{Ball:2012cx}.
The samples are normalised using cross sections computed at
NLO in QCD~\cite{ATL-PHYS-PUB-2016-005}.

The $t$-channel production of a single top quark in association with a $Z$ boson
(\tZ) is generated using \MGAMC\ interfaced with \PYTHIA v6.427~\cite{PythiaManual}, 
referred to as \PYTHIA 6 in the following, 
with the \pdf{CTEQ6L1} PDF~\cite{cteq6} set and the \tune{Perugia2012}~\cite{Skands:2010ak} 
set of tuned MC parameters at NLO in QCD. The \Zgamstar interference is
included, and the four-flavour scheme is used in the computation. 

The \Wt-channel production of a single top quark together with a $Z$ boson
(\WtZ) is generated with  \MGAMC\ and showered with \PYTHIA 8, using the 
\pdf{NNPDF3\!.\!0NLO} PDF set~\cite{Ball:2015cx} and the \tune{A14} tune. The generation is performed at
NLO in QCD using the five-flavour scheme. 
Diagrams containing a top-quark pair are removed to avoid 
overlap with the \ttZ process.

Diboson processes with four charged leptons ($4\ell$), three charged leptons and
one neutrino ($\ell\ell\ell\nu$) or two charged leptons and two neutrinos
($\ell\ell\nu\nu$) are simulated using the \SHERPA 2.1
generator~\cite{Gleisberg:2008ta}. 
The matrix elements include all diagrams with four electroweak vertices. 
They are calculated for up to one ($4\ell,
\ell\ell\nu\nu$) or no additional partons ($\ell\ell\ell\nu$) at NLO and up to three
partons at LO using the \COMIX~\cite{Gleisberg:2008fv} and
\OPENLOOPS~\cite{Cascioli:2011va} matrix element generators and merged with the
\SHERPA parton shower using the \textsc{ME+PS@NLO}
prescription~\cite{Hoeche:2012yf}. The \pdf{CT10nlo} PDF set~\cite{Lai:2010vv} is used in conjunction
with a dedicated parton-shower tuning developed by the \SHERPA authors. The NLO
cross sections calculated by the generator are used to normalise diboson
processes.  Alternative diboson samples are simulated using the \POWHEGBOX
v2~\cite{Powbox2} generator, interfaced to the \PYTHIA 8 parton shower model,
and for which the \pdf{CT10nlo} PDF set is used in the matrix element, while the
\pdf{CTEQ6L1} PDF set is used for the parton shower along with the
\tune{AZNLO}~\cite{AZNLO:2014} set of tuned MC parameters.

The production of three massive vector bosons with subsequent leptonic decays of
all three bosons is modelled at LO with the \SHERPA 2.1 generator and the
\pdf{CT10} PDF set~\cite{Lai:2010vv}. Up to two additional partons are included in the matrix
element at LO and the full NLO accuracy is used for the inclusive process.  

Electroweak processes involving the vector-boson scattering (VBS) diagram and
producing two same-sign leptons,  two neutrinos and two partons are modelled
using \SHERPA 2.1 at LO accuracy and the \pdf{CT10} PDF set. Processes of
orders four and six in the electroweak coupling constant are considered, and up
to one additional parton is included in the matrix element. 

For the generation of \ttbar events and \Wt-channel single-top-quark events the
\POWHEGBOX v2 generator is used with the \pdf{CT10} PDF set. The parton shower
and the underlying event are simulated using \PYTHIA 6
with the \pdf{CTEQ6L1} PDF set and the corresponding
\tune{Perugia2012} tune.  The \ttbar samples are normalised to their
next-to-next-to-leading-order (NNLO) cross-section predictions, including soft-gluon
resummation to next-to-next-to-leading-log order, as calculated with the
\tool{Top++2.0} program (see Ref.~\cite{ref:xs6} and references therein).  For more
efficient sample generation, the \ttbar sample is produced by selecting only true
dilepton events in the final state.  Moreover, an additional dilepton \ttbar
sample requiring a $b$-hadron not coming from top-quark decays is generated
after $b$-jet selection. Diagram removal is employed to remove
the overlap between \ttbar and \Wt~\cite{Re:2010bp}.

Samples of \ttbar events produced in association with a Higgs boson (\ttH) 
are generated using NLO matrix elements in \MGAMC\ with
the \pdf{CT10NLO} PDF set and interfaced with \PYTHIA 8 for the modelling of the
parton shower.  Higgs boson production via gluon--gluon fusion (ggF) and vector
boson fusion (VBF) is generated using the \POWHEGBOX v2 generator with
\pdf{CT10} PDF set. The parton shower and underlying event are simulated using
\PYTHIA 8 with the \pdf{CTEQ6L1} PDF set and \tune{AZNLO} tune.  Higgs boson
production with a vector boson is generated at LO using \PYTHIA 8 with the
\pdf{CTEQ6L1} PDF.  All Higgs boson samples are normalised using theoretical
calculations of Ref.~\cite{lhcxs}. 

Events containing $Z$ or $W$ bosons with associated jets, referred to as 
$Z$+jets and $W$+jets in the following, are simulated using the
\SHERPA 2.1 generator. Matrix elements are calculated for up to two partons at
NLO and four partons at LO.  The \pdf{CT10} PDF set is used in conjunction with
a dedicated parton-shower tuning developed by the \SHERPA authors~\cite{Gleisberg:2008ta}.  The
$Z/W$+jets samples are normalised to the NNLO cross
sections~\cite{Anastasiou:2003ds, Gavin:2010az, Gavin:2012sy, Li:2012wna}.
Alternative $Z/W$+jets samples
are simulated using \MGAMC\ at LO interfaced to the \PYTHIA 8 parton shower
model. The \tune{A14} tune is used together with the \pdf{NNPDF2.3LO} PDF set. 

The SM production of three and four top quarks is generated at LO with
\MGAMC+\PYTHIA 8, using the \tune{A14} tune together with the
\pdf{NNPDF2\!.\!3LO} PDF set. The samples are normalised using cross
sections computed at NLO~\cite{Barger:2010uw,Bevilacqua:2012em}.

\section{Object reconstruction}
\label{s:objetcs}

The final states of interest in this analysis contain electrons, muons, jets,
$b$-jets and missing transverse momentum.

Electron candidates~\cite{PERF-2013-03} are reconstructed from energy deposits
(clusters) in the EM calorimeter that are associated with reconstructed tracks
in the inner detector.  The electron identification relies on a
likelihood-based selection~\cite{ATLAS-CONF-2014-032,
ATL-PHYS-PUB-2015-041}. Electrons are required to pass the `medium'
likelihood identification requirements described in Ref.~\cite{ATL-PHYS-PUB-2015-041}. These include requirements on the 
shapes of the electromagnetic shower in the calorimeter as well as tracking and track-to-cluster matching quantities. 
The electrons are also required to have transverse momentum $\pt > 7\,\GeV$ and
$|\eta_\text{cluster}| < 2.47$, where $\eta_\text{cluster}$ is the
pseudorapidity of the calorimeter energy deposit associated with the electron
candidate.  Candidates in the EM calorimeter barrel/endcap transition region
$1.37 < |\eta_\text{cluster}| < 1.52$ are excluded.  

Muon candidates are reconstructed from a fit to track segments in the various layers of
the muon spectro\-meter, matched with tracks identified in the inner detector.
Muons are required to have $\pt > 7\,\GeV$ and $\abseta < 2.4$ and to pass the
`medium' identification requirements defined in Ref.~\cite{PERF-2015-10}. The `medium' requirement includes selections on the numbers of hits
in the ID and MS as well as a compatibility requirement between momentum measurements in the ID and MS. It provides a high efficiency and purity of selected muons. Electron candidates sharing a track with a muon candidate are removed.  

To reduce the non-prompt lepton background from hadron decays or jets
misidentified as leptons (labelled as ``fake leptons'' throughout this paper), electron and muon
candidates are required to be isolated. The total sum of track transverse
momenta in a surrounding cone of size $\min(10\,\GeV/\pT, r_{e,\mu})$, excluding the track of the candidate from the sum, is required to be less than 6\% of the candidate \pt, where
$r_e = 0.2$ and $r_{\mu} = 0.3$.  In addition, the sum of the cluster
transverse energies in the calorimeter within a cone of size $\Delta R_{\eta} \equiv
\sqrt{(\Delta\eta)^2 + (\Delta\phi)^2} = 0.2$ of any electron candidate,
excluding energy deposits of the candidate itself, is required to be less than
6\% of the candidate \pt.

For both electrons and muons, the longitudinal impact parameter of the
associated track with respect to the primary vertex,\footnote{A primary vertex
candidate is defined as a vertex with at least five associated tracks,
consistent with the beam collision region.  If more than one such vertex is
found, the vertex candidate with the largest sum of squared transverse momenta
of its associated tracks is taken as the primary vertex.} $z_{0}$, is required
to satisfy $|z_0 \sin\theta|<0.5$ mm. The significance of the transverse impact
parameter $d_0$ is required to satisfy $|d_0|/\sigma(d_0)<5$ for electrons and
$|d_0|/\sigma(d_0)<3$ for muons, where $\sigma(d_0)$ is the uncertainty in
$d_0$.

Jets are reconstructed using the anti-$k_t$ algorithm~\cite{Cacciari:2008gp,
Cacciari:2005hq} with radius parameter $R = 0.4$, starting from topological
clusters in the calorimeters~\cite{Aad:2016upy}. The effect of pile-up on jet
energies is accounted for by a jet-area-based correction~\cite{Cacciari:2008gn}
and the energy resolution of the jets is improved by using global sequential
corrections~\cite{ATLAS-CONF-2015-002}. Jets are calibrated to the hadronic
energy scale using $E$- and $\eta$-dependent calibration factors based on MC
simulations, with in-situ corrections based on Run-1 data~\cite{PERF-2011-03,
ATLAS-CONF-2015-037} and checked with early Run-2
data~\cite{ATL-PHYS-PUB-2015-015}.  Jets are accepted if they fulfil the
requirements $\pT > 25\,\GeV$ and $|\eta| < 2.5$. To reduce the contribution
from jets associated with pile-up, jets with $\pT < 60\,\GeV$ and $|\eta| < 2.4$
are required to satisfy pile-up rejection criteria (JVT), based on a multivariate
combination of track-based variables ~\cite{PERF-2014-03}.


Jets are $b$-tagged as likely to contain $b$-hadrons using the \tool{MV2c20}
algorithm, a multivariate discriminant making use of the long lifetime, large
decay multiplicity, hard fragmentation and high mass of $b$-hadrons
\cite{PERF-2012-04}.  The average efficiency to correctly tag a $b$-jet is
approximately $77\%$, as determined in simulated \ttbar events, but it varies as
a function of \pT and $\eta$.  In simulation, the tagging algorithm gives a
rejection factor of about $130$ against light-quark and gluon jets, and about
$4.5$ against jets containing charm quarks~\cite{ATL-PHYS-PUB-2015-022}.  The
efficiency of $b$-tagging in simulation is corrected to that in data using a
\ttbar-based calibration using Run-1 data~\cite{ATLAS-CONF-2014-004} and
validated with Run-2 data~\cite{ATL-PHYS-PUB-2015-039}.

The missing transverse momentum $\mathbf{p}^\text{miss}_\text{T}$, with
magnitude \met, is a measure of the transverse momentum imbalance due to
particles escaping detection. It is computed~\cite{PERF-2011-07} as the
negative sum of the transverse momenta of all electrons, muons and jets and an
additional soft term. The soft term is constructed from all tracks that are associated 
with the primary vertex but not with any physics object. In this way, the \met is adjusted for the best calibration of the jets
and the other identified physics objects above, while maintaining pile-up
independence in the soft term~\cite{ATL-PHYS-PUB-2015-027,
ATL-PHYS-PUB-2015-023}.

To prevent double-counting of electron energy deposits as jets, the closest jet
within $\Delta R_y = 0.2$ of a reconstructed electron is removed, where $\Delta
R_y \equiv \sqrt{(\Delta y)^2 + (\Delta\phi)^2}$.  If the nearest jet surviving
the above selection is within $\Delta R_y = 0.4$ of an electron, the electron
is discarded to ensure that selected electrons are sufficiently separated from
nearby jet activity.  To reduce the background from muons originating from heavy-flavour particle
decays inside jets, muons are removed if they are separated from the nearest
jet by $\Delta R_y < 0.4$.  However, if this jet has fewer than three
associated tracks, the muon is kept and the jet is removed instead; this avoids
an inefficiency for high-energy muons undergoing significant energy loss in the
calorimeter.

\section{Event selection and background estimation}
\label{s:selection}

Only events collected using single-electron or single-muon triggers are
accepted. The trigger thresholds,  $\pT > 24\,\GeV$ for electrons and
 $\pT > 20\,\GeV$ for muons, are set to be almost fully efficient for
reconstructed leptons with $\pT > 25\,\GeV$.  Events are required
to have at least one reconstructed primary vertex.  In all selections
considered, at least one reconstructed lepton with $\pT > 25\,\GeV$ is required
to match ($\Delta R_{\eta} < 0.15$) a lepton with the same flavour
reconstructed by the trigger algorithm.  Three channels are defined based on
the number of reconstructed leptons, which are sorted according to their
transverse momentum in decreasing order. 

Background events containing well-identified prompt leptons are modelled by simulation. The
normalisations for the $WZ$ and $ZZ$ processes are taken from data control regions and 
included in the fit.
The yields in these data control regions are 
extrapolated to the signal regions using simulation.
Systematic uncertainties in the extrapolation are taken into account in the
overall uncertainty in the background estimate.

Background sources involving one or more fake leptons are modelled 
using data events from control regions.  For the same-sign dimuon (\SSLSR) analysis and the
trilepton analysis the fake-lepton background is estimated using the matrix
method~\cite{TOPQ-2010-01}, where any combination of fake leptons among the
selected leptons is considered.  However, compared to Ref.~\cite{TOPQ-2010-01}, the 
real- and fake-lepton efficiencies used by the matrix method are estimated
in a different way in this measurement. The lepton efficiencies are measured by
applying the matrix method in control regions, where the lepton efficiencies
are extracted in a likelihood fit as free parameters using the matrix method
as model, assuming Poisson statistics, and assuming that events with two fake
leptons are negligible. In this way the parameters are by construction the
actual parameters of the matrix model itself, instead of relying on external
lepton efficiency measurements which are not guaranteed to be fully consistent
with the matrix model.
The
control regions are defined in dilepton events, separately for $b$-tagged and
$b$-vetoed events to take into account the different fake-lepton efficiencies
depending on whether the source is a light-flavour jet or a heavy-flavour jet.
The real-lepton efficiencies are measured in inclusive opposite-sign events,
and fake-lepton efficiencies in events with same-sign leptons and $\met>40\,\GeV$
(for $b$-tagged events $\met>20\,\GeV$), after subtracting the estimated
contribution from events with misidentification of the charge of a lepton
(referred to as ``charge-flip'' in the following), and excluding the same-sign dimuon signal region. The charge-flip events are subtracted using simulation. The
extracted fake-lepton efficiencies are found to be compatible with fake-lepton
efficiencies from a fully data-driven procedure where the charge-flip events
are estimated from data.  For the \FLC, the contribution from backgrounds
containing fake leptons is estimated from simulation and corrected with scale
factors determined in control regions.

The full selection requirements and the background evaluation strategies in the
different channels are described below.

\subsection{Same-sign dimuon analysis}

The same-sign dimuon signal region targets the \ttW process and has
the highest sensitivity among all same-sign dilepton
regions~\cite{TOPQ-2013-05}.  The main reason for this is that electrons have a
much larger charge misidentification probability, inducing a significant
background from top-quark pairs.  Events are required to have two muon
candidates with the same charge and $\pt>25\,\GeV$, $\met > 40\,\GeV$, the scalar
sum of the \pT of selected leptons and jets, \HT, above $240\,\GeV$, and at least
two $b$-tagged jets. Events containing additional leptons (with $\pT>7\,\GeV$)
are vetoed.

\begin{figure}[htbp]
\centering
\includegraphics[width=\twofigwidth]{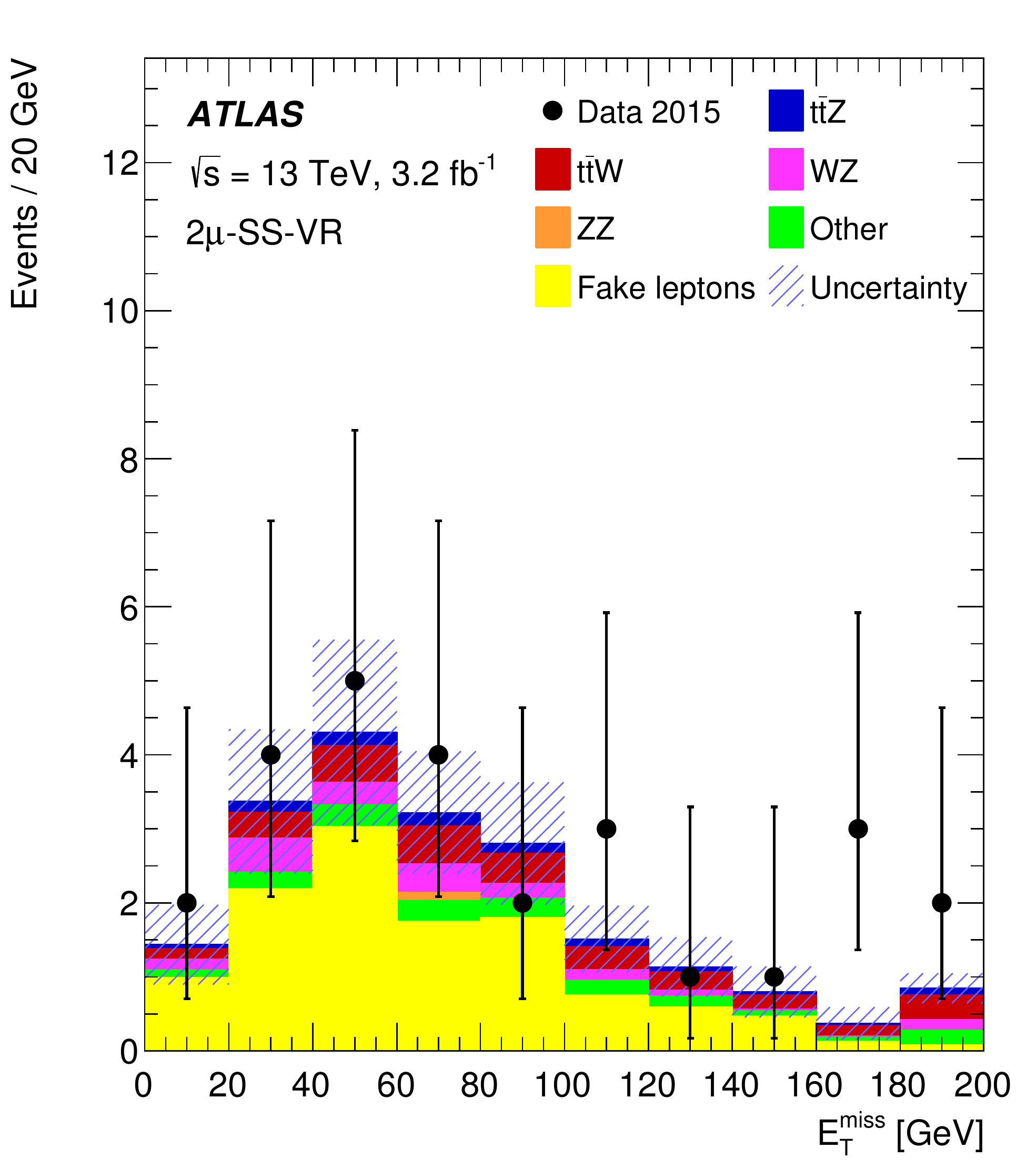}
\includegraphics[width=\twofigwidth]{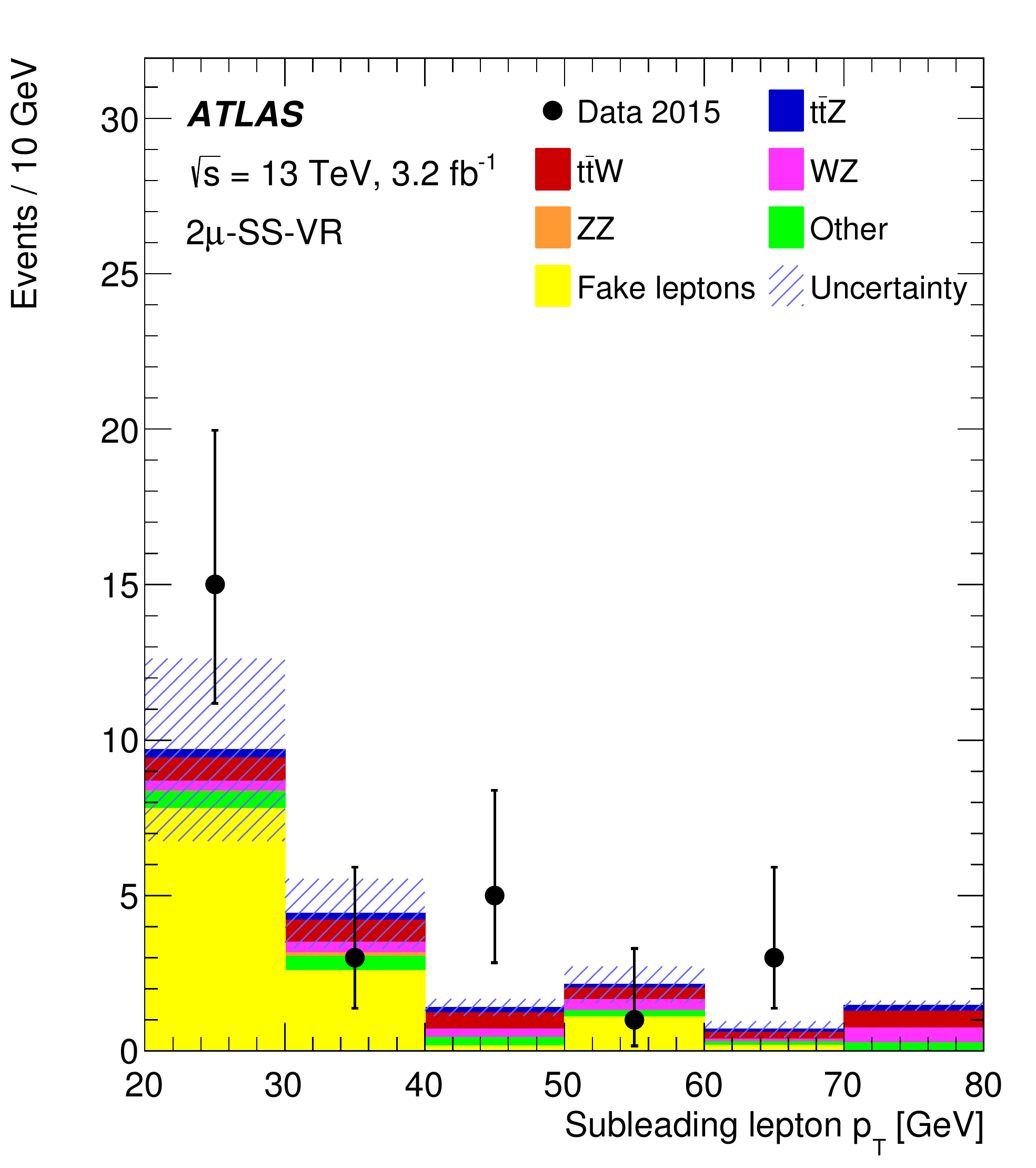}
\caption{\label{fig:ssmm_val} (Left) The \met and (right) subleading lepton \pT 
distributions shown for the $b$-tagged \SSLSR\ channel where the signal region
requirements on subleading lepton \pT, number of $b$-tags, and \met are
relaxed. \hatch.  The background denoted `Other' contains other SM processes
producing two same-sign prompt leptons.  \oflo.} 
\end{figure}

The dominant background in the \SSLSR\ region arises from events containing
fake leptons, where the main source is \ttbar events. Backgrounds from the
production of prompt leptons with correctly
identified charge come primarily from $WZ$ production, but the relative
contribution of this background is small compared to the fake-lepton
background. The charge-flip background is negligible in this signal region, as
the probability of misidentifying the charge of a muon in the relevant \pT
range is negligible.  
For the validation of the fake-lepton background estimate a region is defined
based on the signal region selection but omitting the \met requirement,
reducing the \pT threshold of the subleading lepton to $20\,\GeV$ and requiring 
at least one $b$-tagged jet.
The distributions of \met and subleading lepton \pT in this validation region ($2\mu$-SS-VR) are shown in
Figure \ref{fig:ssmm_val}.  The expected numbers of events in the \SSLSR\
signal region are shown in Table~\ref{tab:yields}. Nine events are
observed in data for this signal region.

\subsection{Trilepton analysis}

Four signal regions with exactly three leptons are considered.  The first three
are sensitive to \ttZ; each of these requires an opposite-sign same-flavour
(OSSF) pair of leptons whose invariant mass is within $10\,\GeV$ of the $Z$ boson mass.
The signal regions are categorised by their jet and $b$-jet multiplicities and
have different signal-to-background ratios.  In the \TLSRA\ region, at least
four jets are required, exactly one of which is $b$-tagged. In the \TLSRB\
region, exactly three jets with at least two $b$-tagged jets are required.  In
the \TLSRC\ region, at least four jets are required, of which at least two are
$b$-tagged. 

In the \TLSRD\ region at least two and at most four jets are required, of which
at least two are $b$-tagged, no OSSF lepton pair is allowed in the $Z$
boson mass window, and the sum of the lepton charges must be $\pm$1.  This region
primarily targets the \ttW process but also has a sizeable \ttZ contribution.

The signal region definitions for the \TLC\ are summarised
in Table~\ref{tab:SRs3l}, while the expected numbers of events in the signal
regions are shown in Table~\ref{tab:yields}.  The dominant backgrounds in the
\TLSRA, \TLSRB\ and \TLSRC\ signal regions arise from $Z$+jets production with
a fake lepton, diboson production and the production of a single top quark in
association with a $Z$ boson.

\begin{table}[htbp]
\centering
\caption{Summary of event selections in the trilepton signal regions.\vspace{1ex}} 
\label{tab:SRs3l}
\begin{tabular}{l|c|c|c|c}
\toprule
Variable  & \TLSRA & \TLSRB & \TLSRC & \TLSRD\\
\midrule
Leading lepton & \multicolumn{4}{c}{$\pT>25\,\GeV$} \\
Other leptons & \multicolumn{4}{c}{$\pT>20\,\GeV$} \\
Sum of lepton charges &  \multicolumn{4}{c}{$\pm1$} \\
$Z$-like OSSF pair & \multicolumn{3}{c|}{$|m_{\ell\ell} - m_Z| < 10\,\GeV$} & $|m_{\ell\ell} - m_Z| > 10\,\GeV$\\ 
$n_{\text{jets}}$ & $\ge 4$ & $3$ &  $\ge 4$ & $\ge2$ and $\le4$\\
$n_{b{\text{-jets}}}$ & $1$ & $\ge2$ & $\ge2$ & $\ge2$\\
\bottomrule
\end{tabular}
\end{table}

\begin{figure}[htbp]
\centering
\includegraphics[width=\twofigwidth]{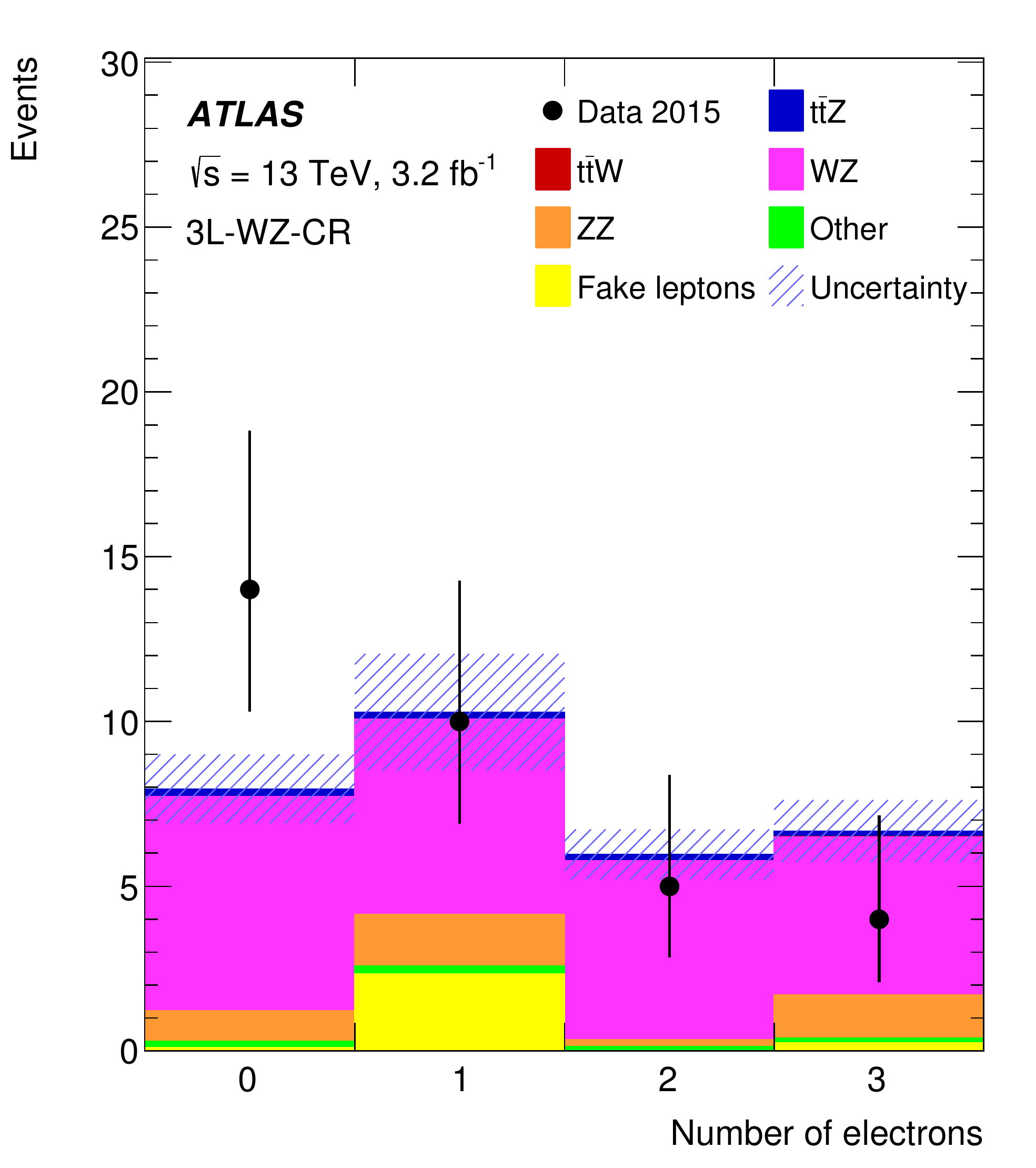}
\includegraphics[width=\twofigwidth]{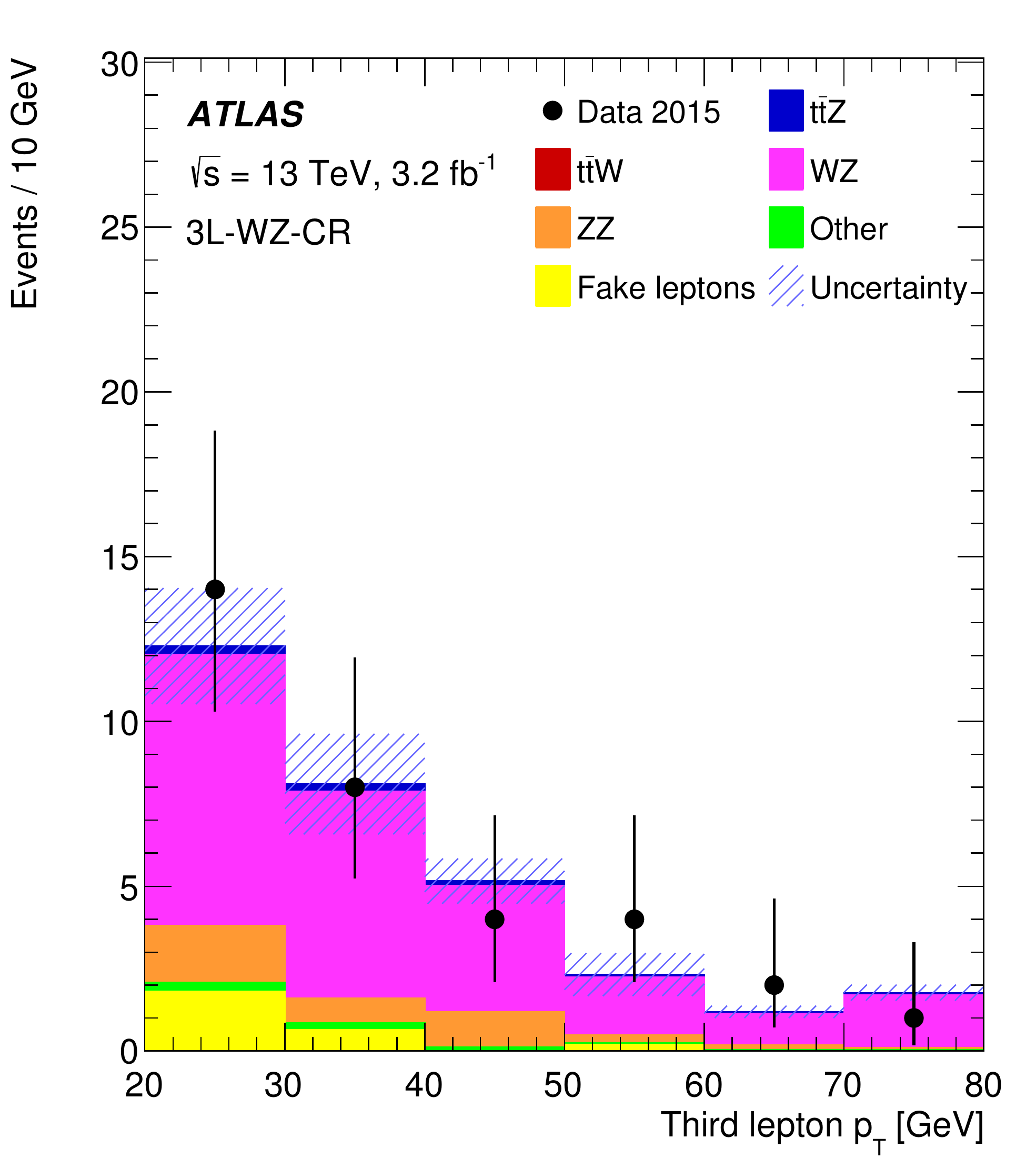}
\caption{\label{fig:3l_wzcr} Distributions of (left) the number of electrons
and (right) the third-lepton \pt in the \TLCR\ control region before the fit.  
The background
denoted `Other' contains other SM processes producing three prompt leptons.
\hatch. \oflor.}
\end{figure}

A control region is used to constrain the normalisation of the $WZ$ background
in data. Exactly three leptons are required, at least one pair of which must 
be an OSSF pair with an invariant mass within $10\,\GeV$ of the $Z$ boson mass. 
There must be exactly three jets,
none of which pass the $b$-tagging requirement.  With these requirements, the
expected \ttZ signal contribution is roughly 1\% of the total number of events.
This region is referred to as \TLCR\ and it is included in the fit. Distributions
comparing data and SM prediction are shown in Figure~\ref{fig:3l_wzcr}.

\begin{figure}[htbp]
\centering
\includegraphics[width=\twofigwidth]{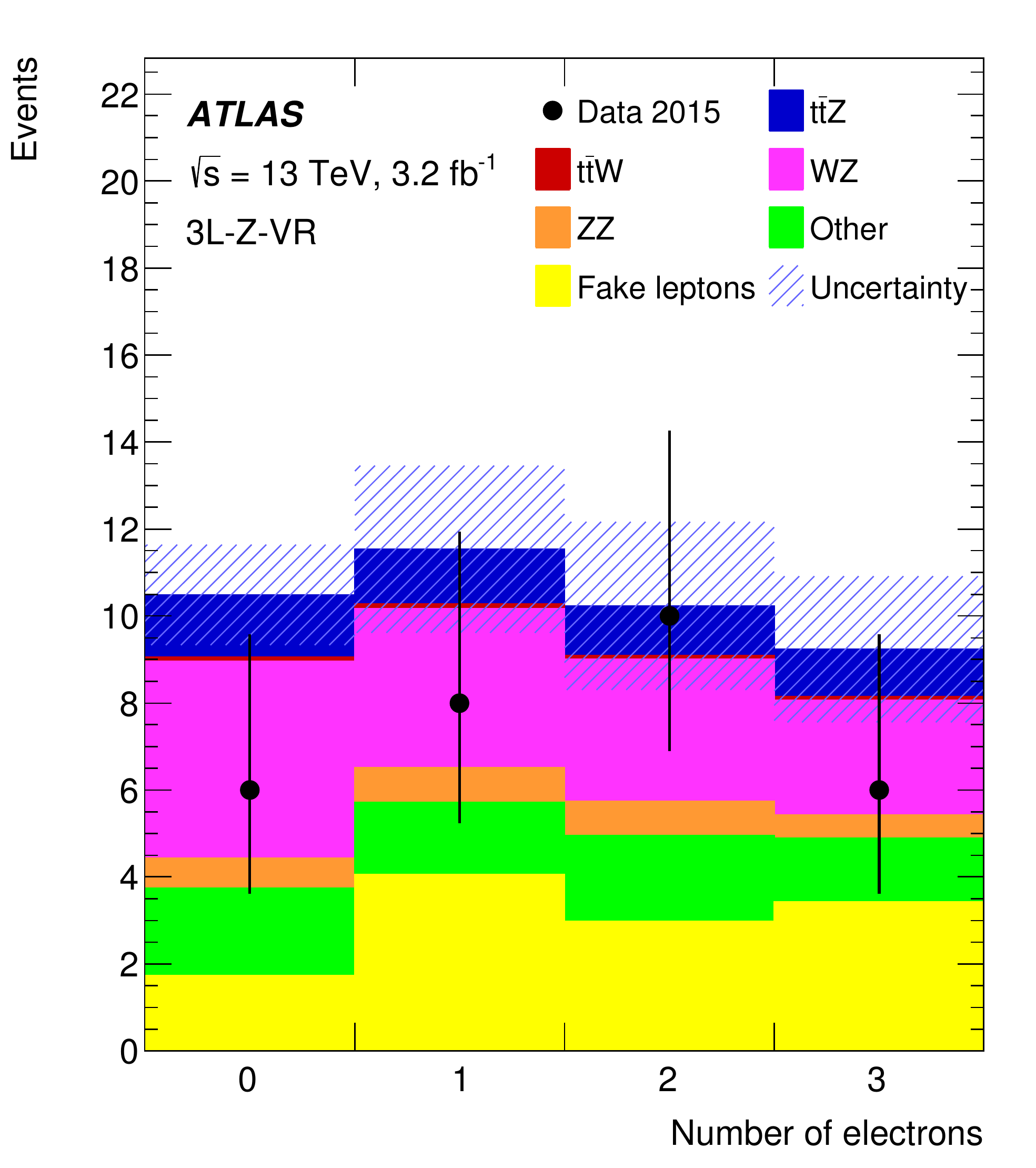}
\includegraphics[width=\twofigwidth]{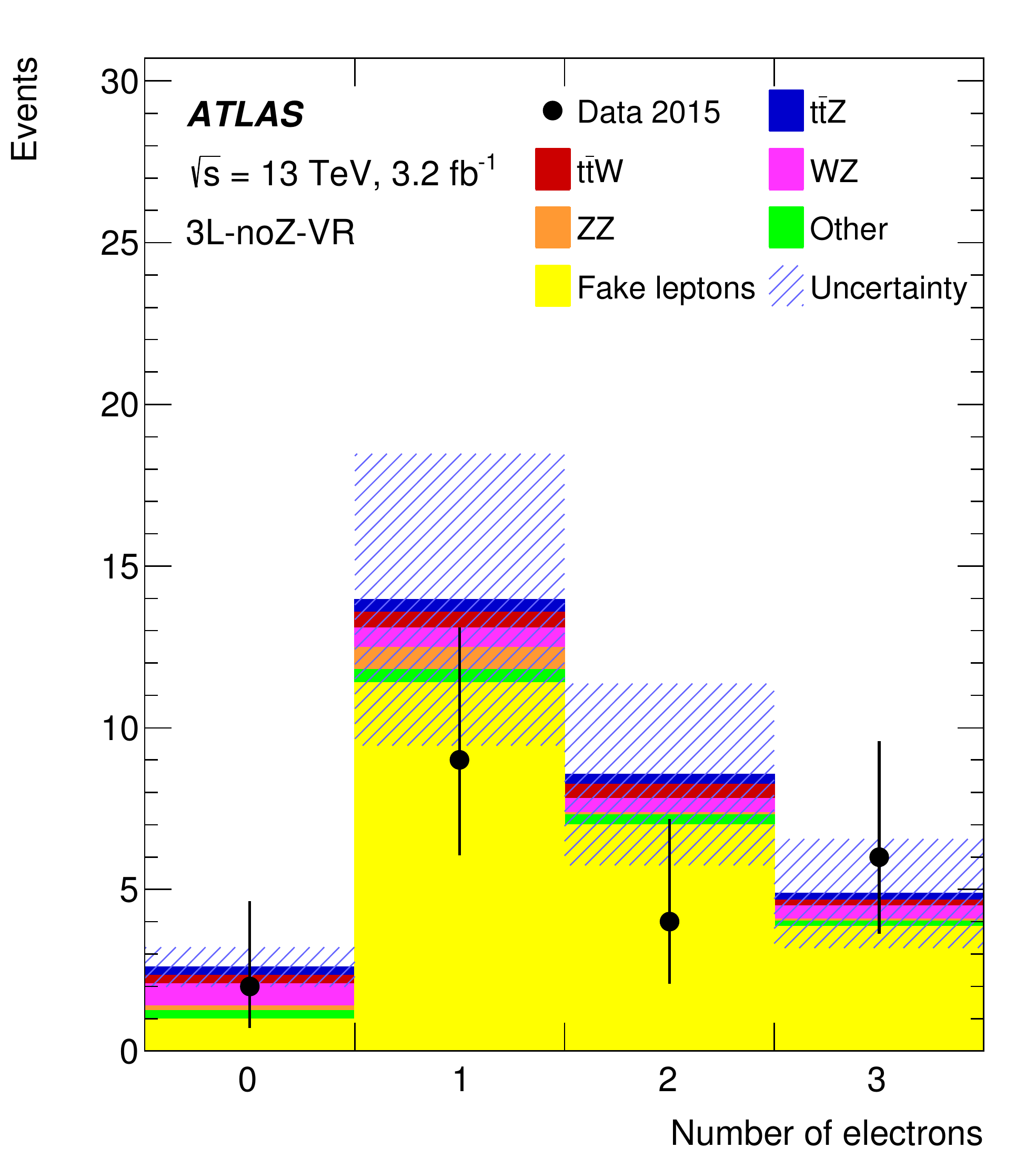}
\caption{\label{fig:3l_val} Distributions of the number of electrons in the
(left) $3\ell$-Z-VR and (right) $3\ell$-noZ-VR validation regions, shown before
the fit.  The background denoted `Other' contains other SM processes producing
three prompt leptons.  \hatch.}
\end{figure}

Two background validation regions are defined for the \TLC.  In the first
region, $3\ell$-Z-VR, the presence of two OSSF leptons with an invariant mass
within $10\,\GeV$ of the mass of the $Z$ boson is required.  The region requires
the events to have at most three jets where exactly one is $b$-tagged, or
exactly two jets where both jets are $b$-tagged.  The main backgrounds are $WZ$
production and $Z$+jets events with fake leptons.  In the second region,
$3\ell$-noZ-VR, events with such a pair of leptons are vetoed.  This region
requires the events to have at most three jets where exactly one is $b$-tagged,
and it is dominated by the fake-lepton background from top-quark pair production.
Neither validation region is used in the fit.  The distributions of the number
of electrons in each of the two validation regions are shown in
Figure~\ref{fig:3l_val}, demonstrating that data and background modelling are 
in good agreement within statistical uncertainties.

\begin{figure}[htbp]
\centering
\includegraphics[width=\twofigwidth]{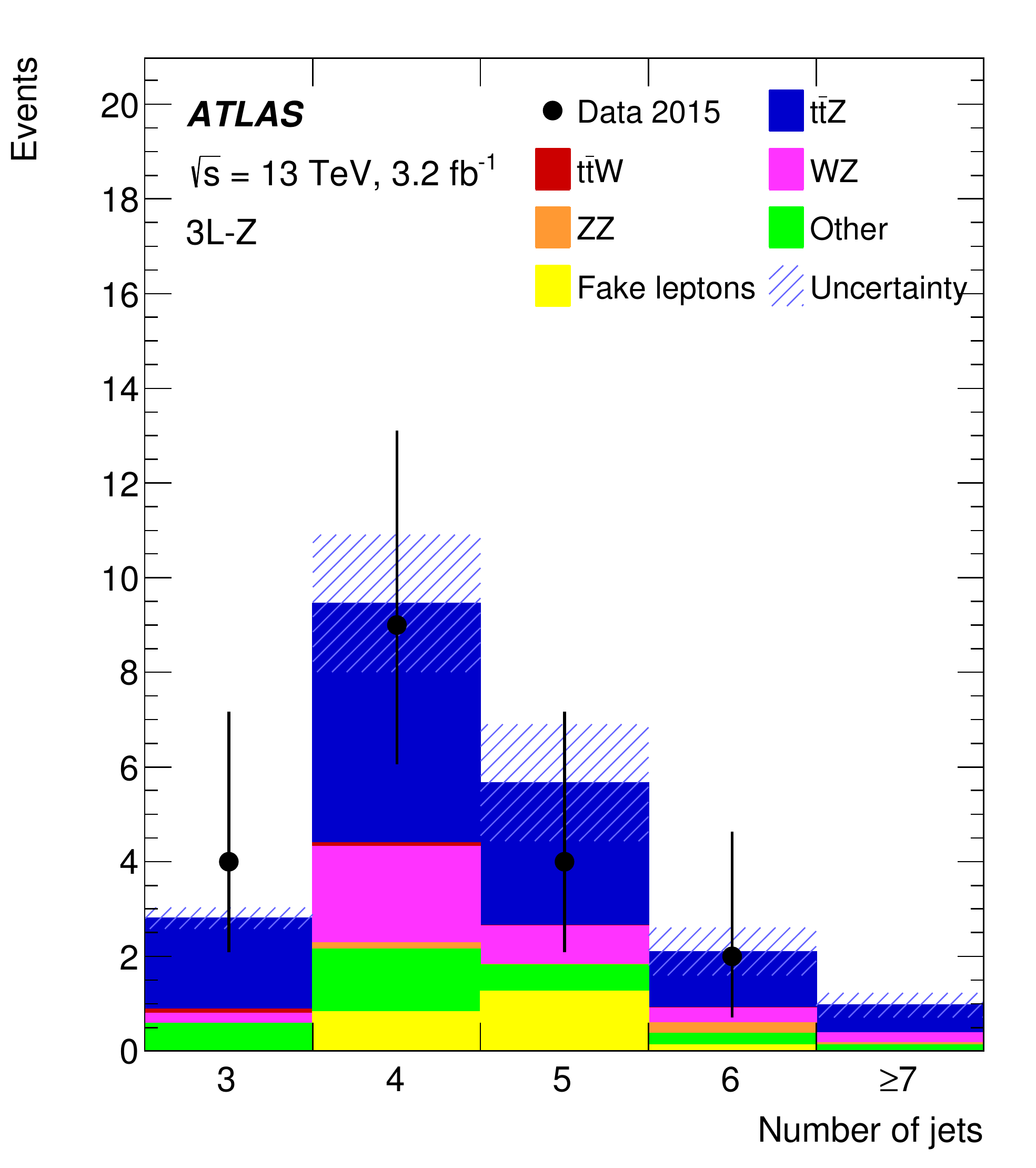}
\includegraphics[width=\twofigwidth]{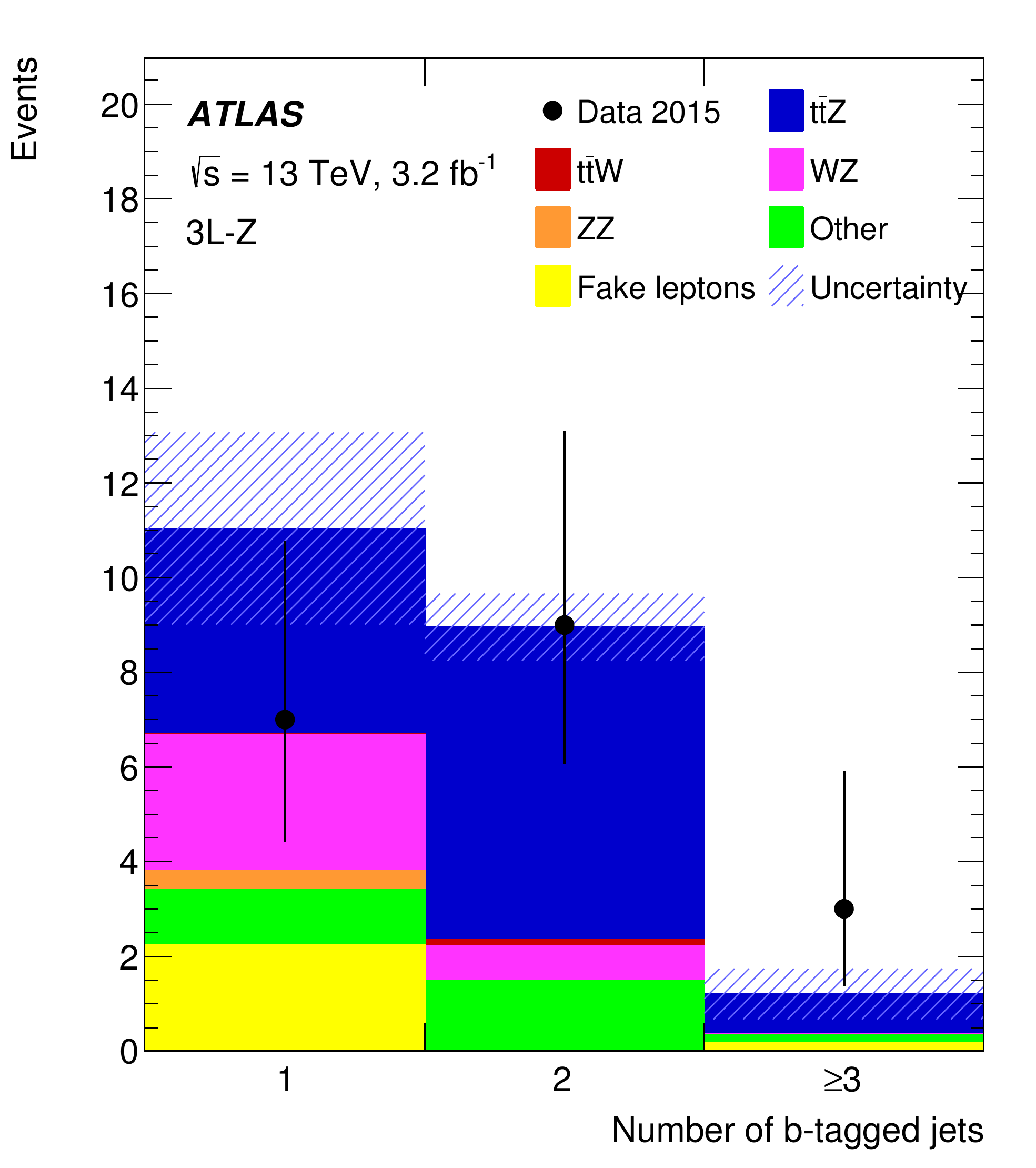}
\includegraphics[width=\twofigwidth]{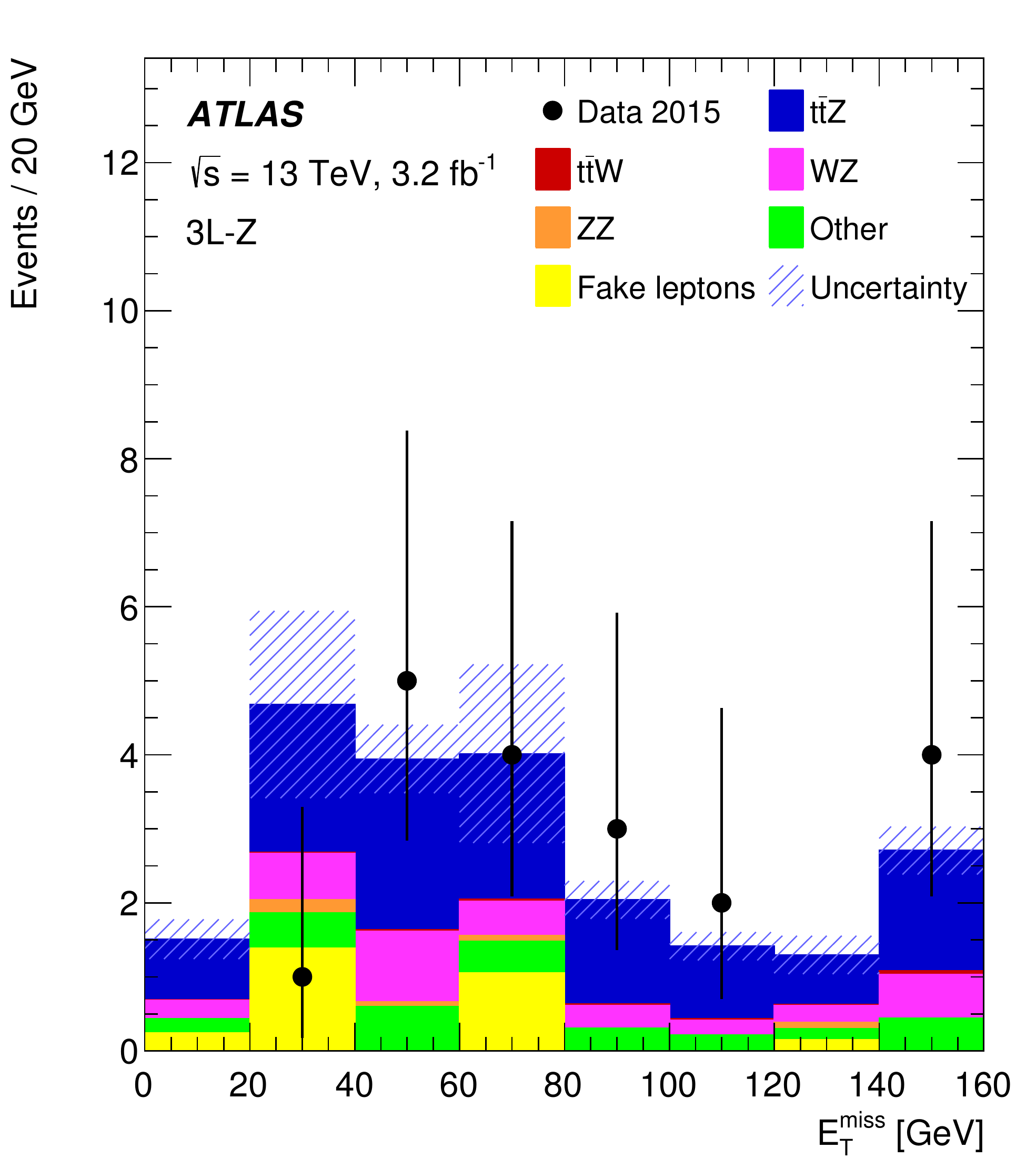}
\includegraphics[width=\twofigwidth]{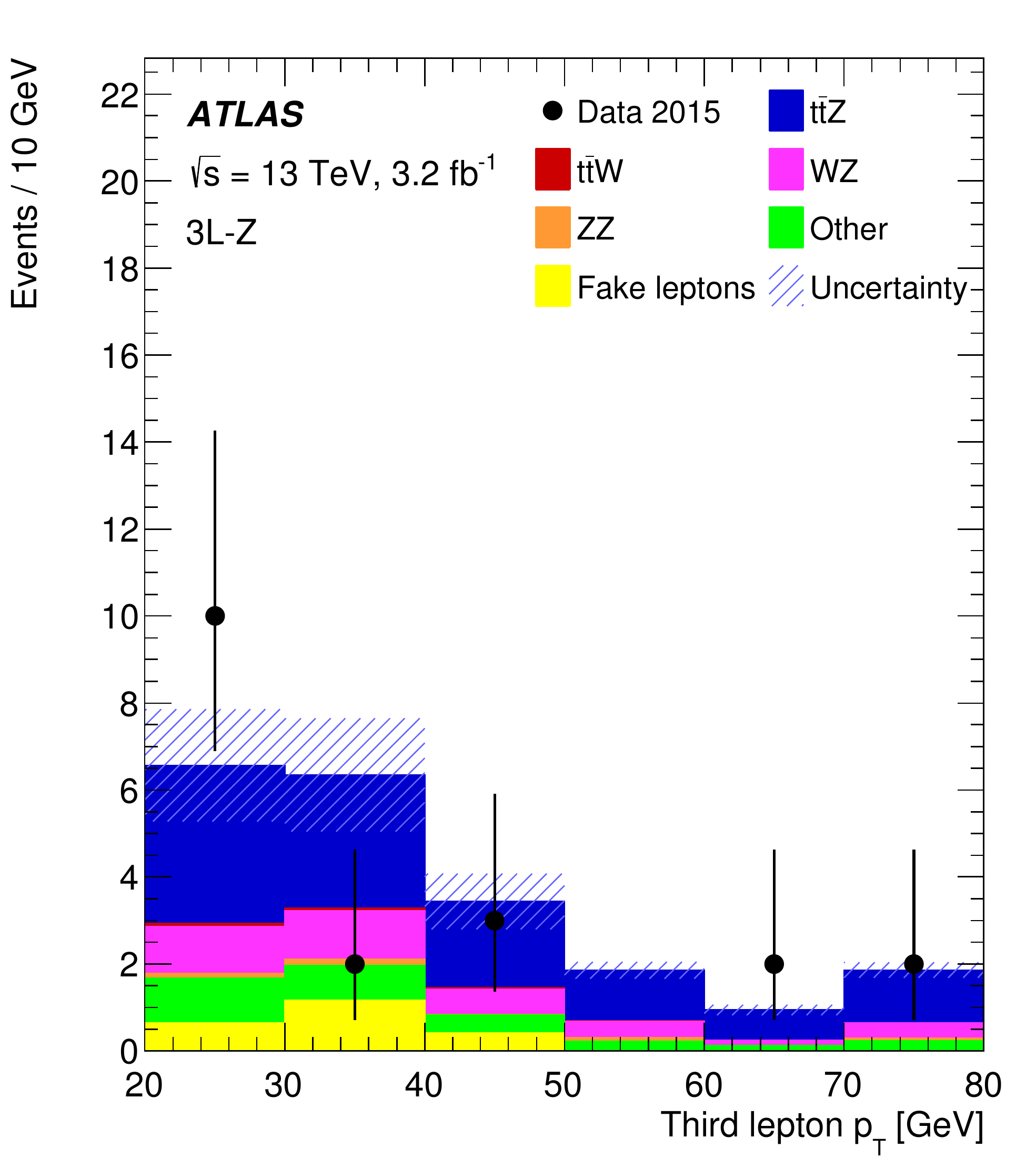}
\caption{\label{fig:3l_sr} Distributions of (top left) the number of jets, (top
right) the number of $b$-tagged jets, (bottom left) the missing transverse
momentum and (bottom right) the third-lepton \pt, for events contained in any
of the three signal regions \TLSRA, \TLSRB\ or \TLSRC. The distributions are
shown before the fit.  The background denoted `Other' contains other SM
processes producing three prompt leptons.  \hatch. The last bin in each of the
distributions shown in the bottom panels includes the overflow.}
\end{figure}

In total, 29 events are observed in the four signal regions. Distributions of
the number of jets, number of $b$-tagged jets, missing transverse momentum and
transverse momentum of the third lepton are shown in Figure~\ref{fig:3l_sr}.

\subsection{Tetralepton analysis}
\label{s:4L}

The \FLC\ targets the \ttZ process for the case where both $W$ bosons resulting
from top-quark decays and the $Z$ boson decay leptonically.  Events with two
pairs of opposite-sign leptons are selected, and at least one pair must be
of same flavour. The OSSF lepton pair with reconstructed invariant mass closest to
$m_Z$ is attributed to the \Zboson boson decay and denoted in the following by
$Z_1$.  The two remaining leptons are used to define $Z_2$.  Four signal
regions are defined according to the relative flavour of the two $Z_2$ leptons,
different flavour (DF) or same flavour (SF), and the number of $b$-tagged jets:
one, or at least two ($1b$, $2b$).  The signal regions are thus \FLSRB, \FLSRC,
\FLSRD\ and \FLSRE.

To suppress events with fake leptons in the 1-$b$-tag multiplicity regions, additional requirements on the scalar sum of the transverse momenta of the
third and fourth leptons ($\pttf$) are imposed.  In the \FLSRD\ and \FLSRB\
regions, events are required to satisfy $\pttf > \SI{25}{\gev}$ and $\pttf >
\SI{35}{\gev}$, respectively.  In all regions, the invariant mass of any two
reconstructed OS leptons is required to be larger than \SI{10}{\gev}. The signal region definitions for the \FLC\ are summarised in Table~\ref{tab:SRs4l}.

\begin{table}[htbp]
\centering \renewcommand{\arraystretch}{1.2}
\caption{\label{tab:SRs4l} Definitions of the four signal regions in the \FLC.
All leptons are required to satisfy \mbox{$\pT > 7\,\GeV$} and at least one lepton with
$\pt > 25\,\GeV$ is required to be trigger matched. The invariant mass of any two
reconstructed OS leptons is required to be larger than \SI{10}{\gev}.\vspace{1ex}}
\begin{tabular}{lccr@{}cc@{}lc}
\toprule  
Region & $Z_2$ leptons &  \pttf && $|m_{Z_{2}} - m_Z| $ & \met && $n_{b{\text{-tags}}}$\\
\midrule
\FLSRB & $e^{\pm}\mu^{\mp}$ & $>\SI{35}{\gev}$ &&-&-&& 1\\
\FLSRC & $e^{\pm}\mu^{\mp}$ & -&&-&-&& $\ge2$\\
\FLSRD & $e^{\pm}e^{\mp},\mu^{\pm}\mu^{\mp}$ & $>\SI{25}{\gev}$ & \begin{tabular}{@{}r@{}} \ldelim\{{2}{2ex} \\ \\ \end{tabular} &
\begin{tabular}{c}$>\SI{10}{\gev}$\\ $<\SI{10}{\gev}$\end{tabular} & \begin{tabular}{c}$>\SI{40}{\gev}$\\ $>\SI{80}{\gev}$\end{tabular} &\begin{tabular}{@{}l@{}} \rdelim\}{2}{2ex} \\ \\ \end{tabular} & 1\\
\FLSRE & $e^{\pm}e^{\mp},\mu^{\pm}\mu^{\mp}$ &-&\begin{tabular}{@{}r@{}} \ldelim\{{2}{2ex} \\ \\ \end{tabular} &
\begin{tabular}{c}$>\SI{10}{\gev}$\\ $<\SI{10}{\gev}$\end{tabular} & \begin{tabular}{c}-\\ $>\SI{40}{\gev}$\end{tabular} &\begin{tabular}{@{}l@{}} \rdelim\}{2}{2ex} \\ \\ \end{tabular} & $\ge2$ \\
\bottomrule 
\end{tabular}
\end{table}

\begin{figure}[htbp]
\centering
\includegraphics[width=\twofigwidth]{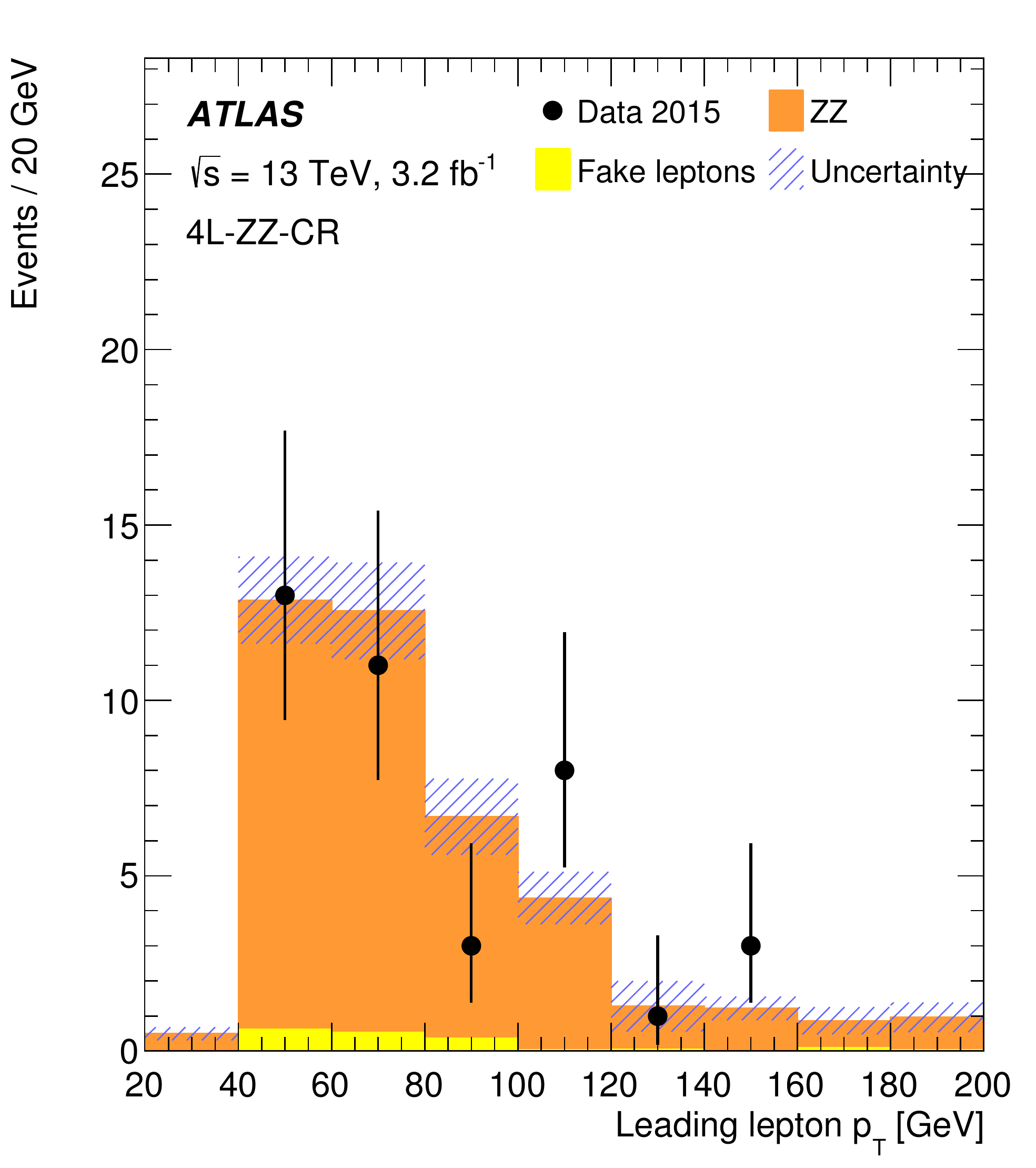}
\includegraphics[width=\twofigwidth]{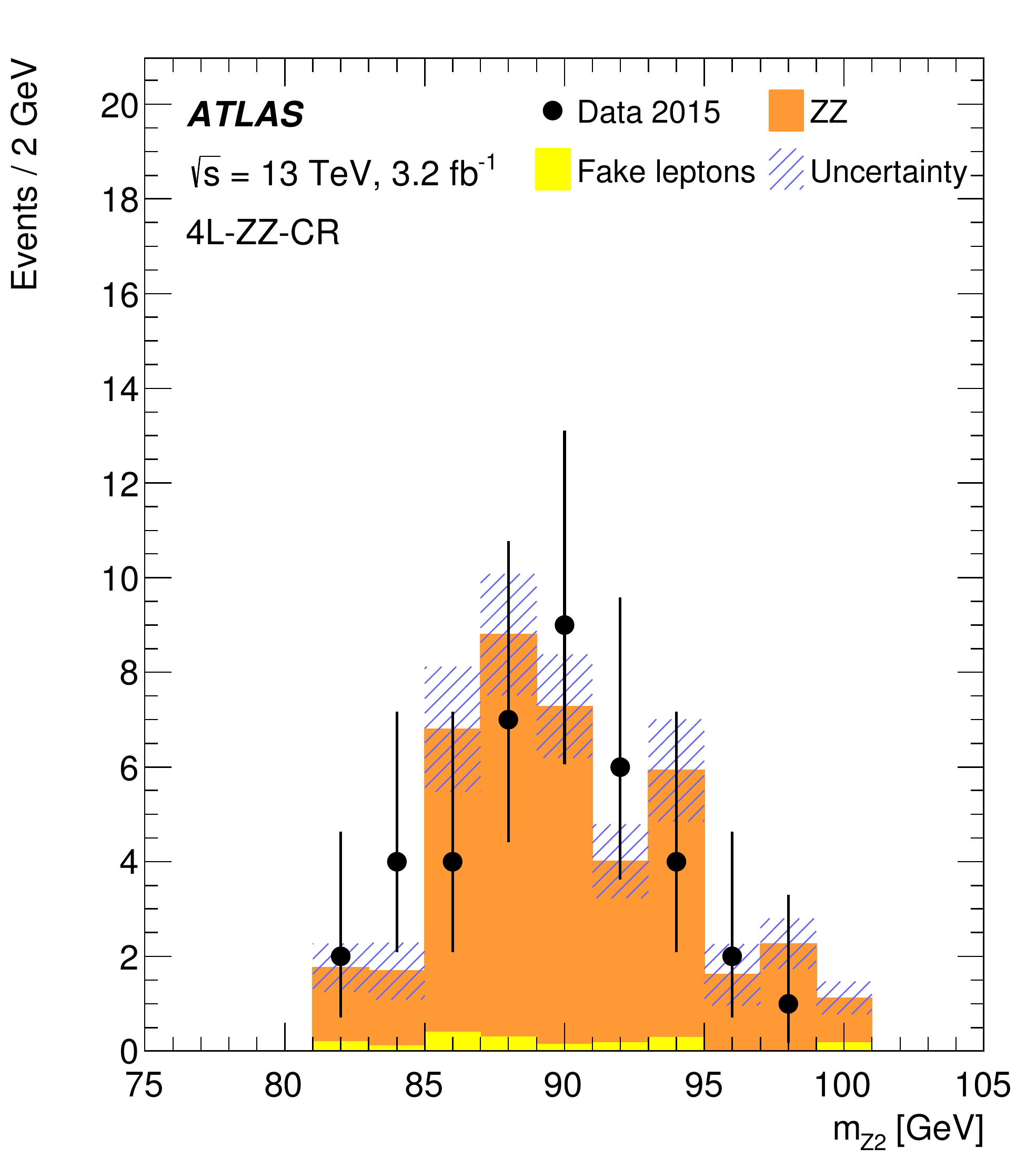}
\includegraphics[width=\twofigwidth]{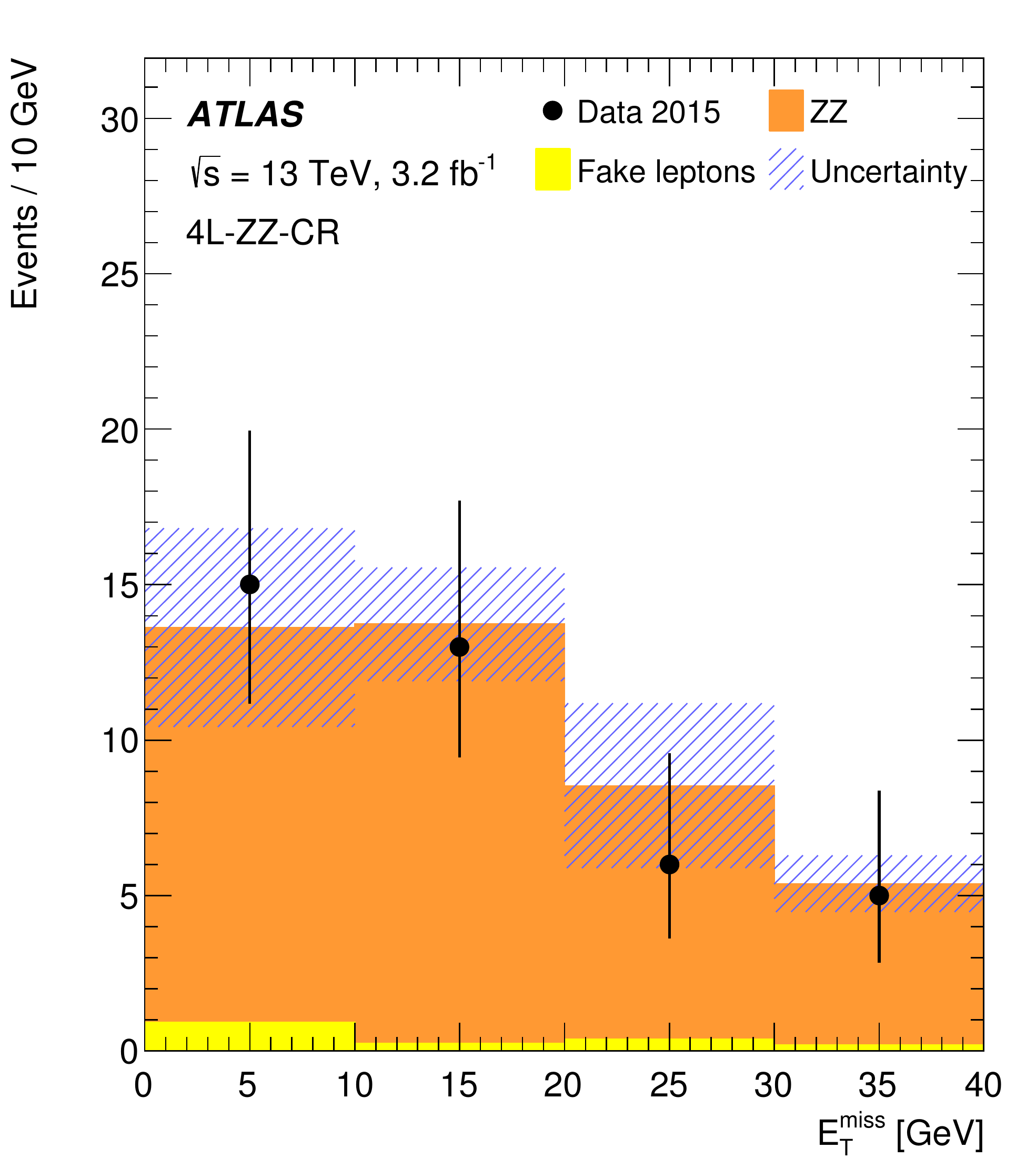}
\includegraphics[width=\twofigwidth]{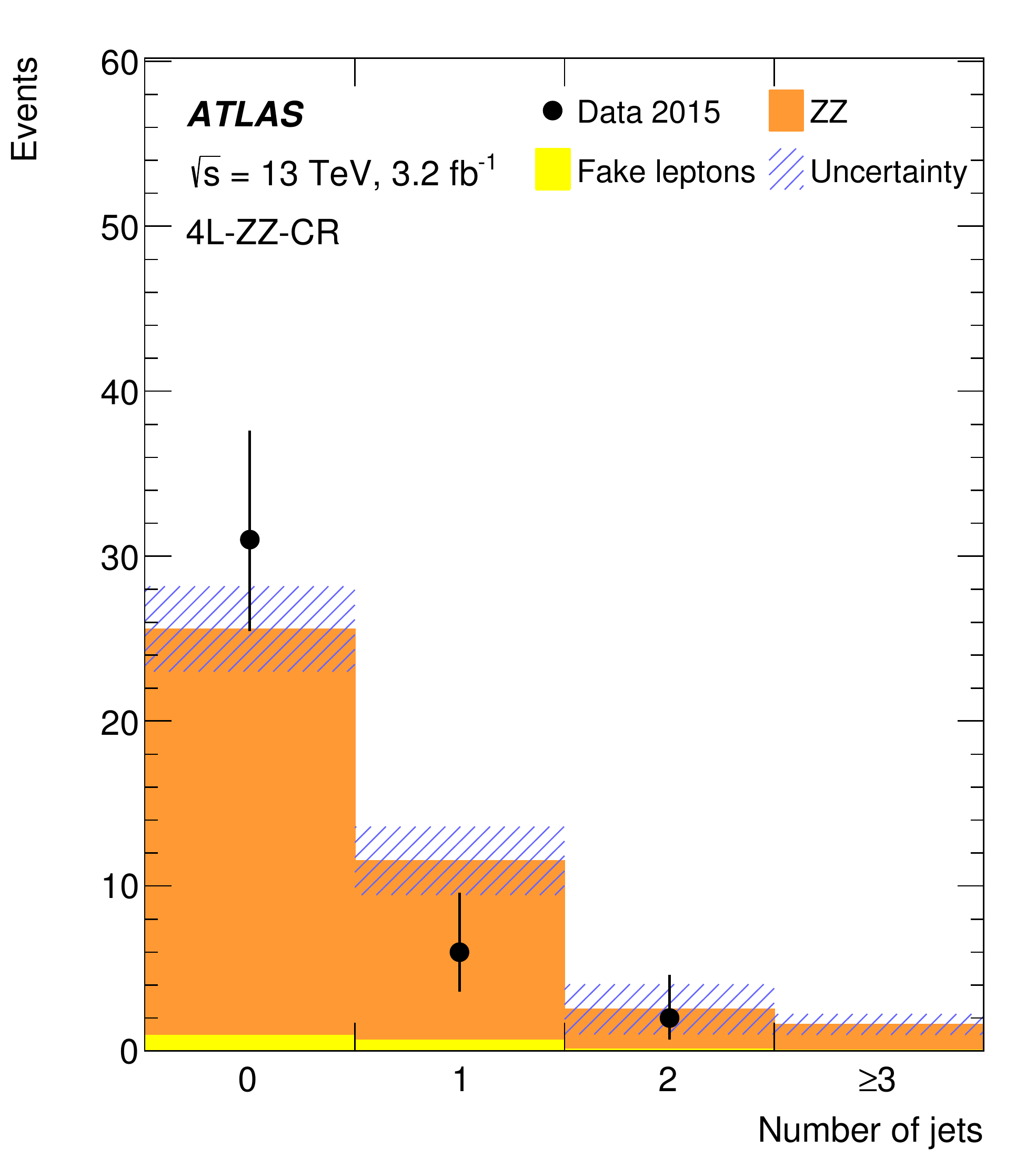}
\caption{\label{fig:zz_val} (Top left) Leading lepton \pt, (top right)
$m_{Z_2}$, (bottom left) missing transverse momentum and (bottom right) jet
multiplicity distributions in the \FLCR\ control region. The distributions are shown
before the fit. \hatch. The last bin of the distribution shown in the top left
panel includes the overflow.} 
\end{figure}

A control region used to constrain the $ZZ$ normalisation, referred to as \FLCR, 
is included in the fit and
is defined to have exactly four reconstructed leptons, a \SecondZ\ pair with
OSSF leptons, the value of both \FirstZM\ and \SecondZM\ within $10\,\GeV$ of the
mass of the $Z$ boson, and $\met <40\,\GeV$. The leading lepton \pT, the 
invariant mass of the $Z_2$ lepton pair, the missing transverse momentum and the 
jet multiplicity in this control region are shown in Figure~\ref{fig:zz_val},
and good agreement is seen between data and prediction.

The contribution from backgrounds containing fake leptons is estimated from
simulation and corrected with scale factors determined in two control regions:
one region enriched in \ttbar events and thus in heavy-flavour jets, and one
region enriched in $Z$+jets events, and thus in light-flavour jets.  The scale factors
are calibrated separately for electron and muon fake-lepton candidates.  The
scale factors are applied to all MC simulation events with fewer than four
prompt leptons according to the number and the flavour of the fake leptons.
The \ttbar scale factors are applied to MC processes with real top quarks,
while for all other processes the $Z$+jets scale factors are applied. Different
generators are used when determining the scale factors and when applying them.
It is verified that the uncertainties in the scale factors include the differences between these generators.

\begin{figure}[htbp]
\centering
\includegraphics[width=\twofigwidth]{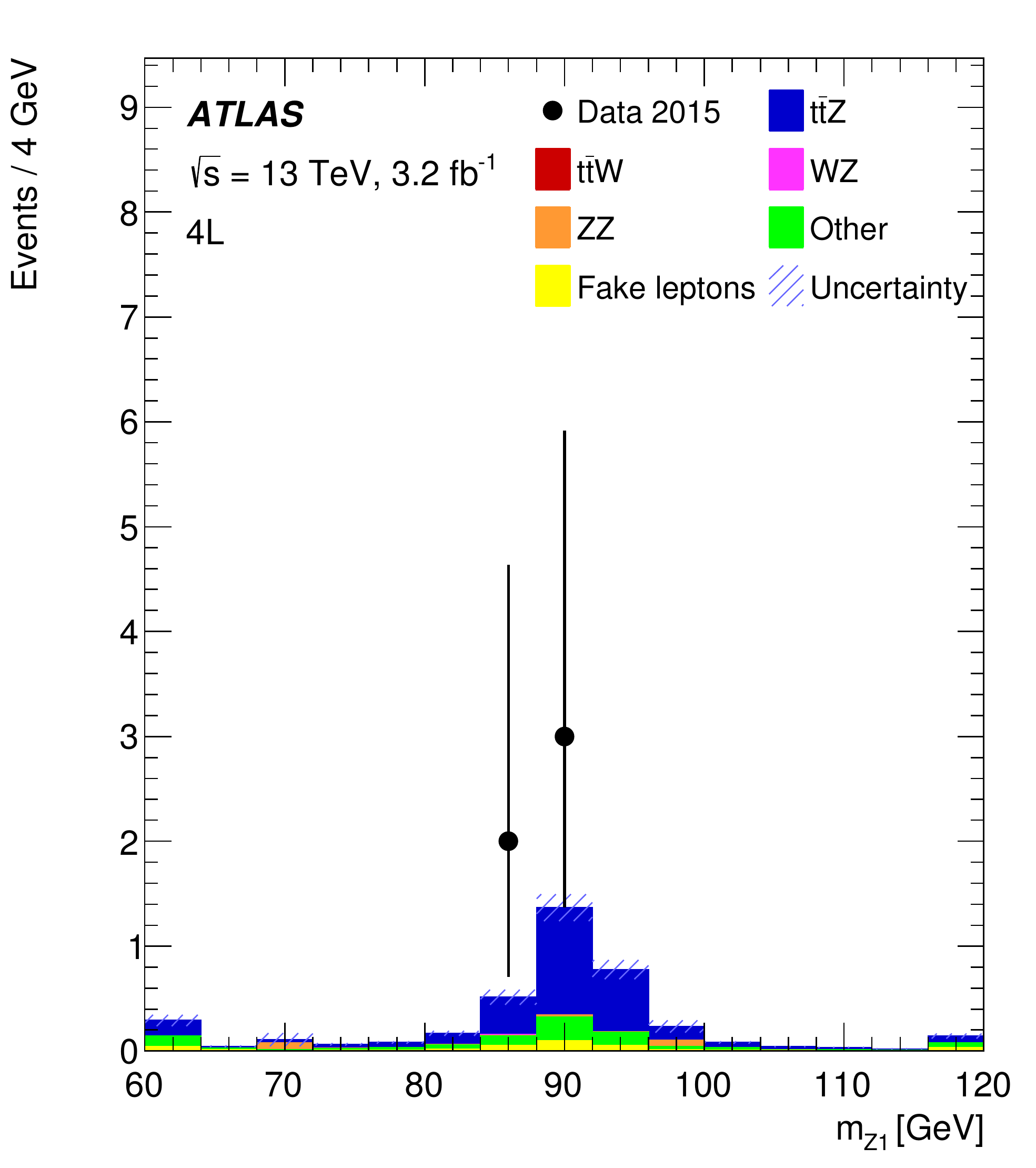}
\includegraphics[width=\twofigwidth]{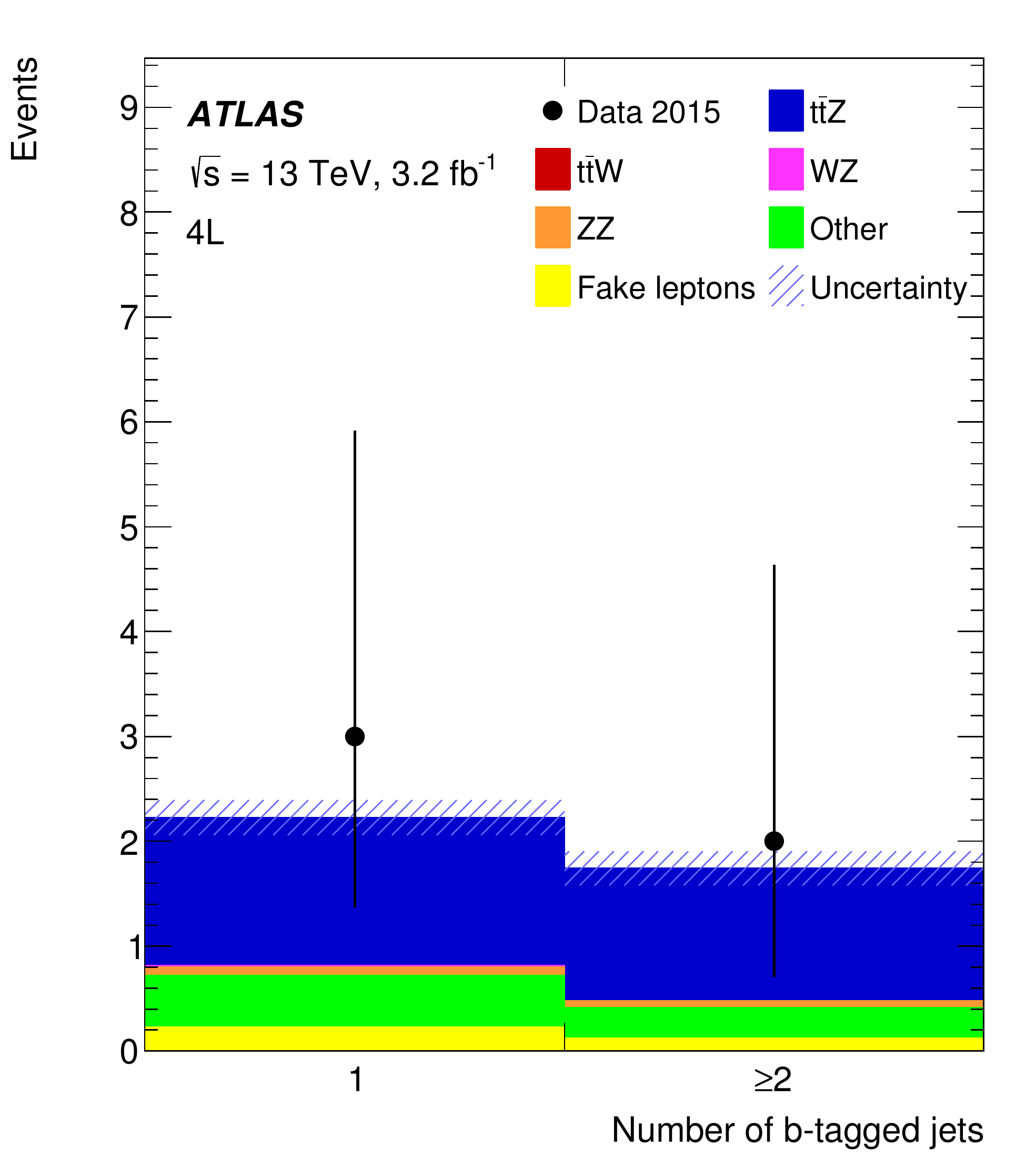}
\caption{\label{fig:SR4l} Distributions (left) of the invariant mass of the
OSSF lepton pair closest to the $Z$ boson mass, $m_{Z_1}$, and (right) of the
number of $b$-tagged jets, for events in the tetralepton signal regions.  The
distributions are shown before the fit.  The background denoted `Other'
contains other SM processes producing four prompt leptons.  \hatch.  The first
and last bin of the distribution shown in the left panel include the underflow
and overflow, respectively.} 
\end{figure}

The expected yields in the signal and control regions in the \FLC\ are shown
in Table~\ref{tab:yields}.  Five events are observed in the four signal
regions. Figure~\ref{fig:SR4l} shows the data superimposed to the expected
distributions for all four signal regions combined.
Overall the acceptance times efficiency for the \ttZ and \ttW
processes is 6\textperthousand\ and 2\textperthousand, respectively.  

\begin{table}[htbp]
\centering \renewcommand{\arraystretch}{1.2}
\caption{\label{tab:yields} Expected event yields for signal and backgrounds,
and the observed data in all control and signal regions used in the fit to
extract the \ttZ and \ttW cross sections.  The quoted uncertainties in the expected
event yields represent systematic uncertainties including MC statistical
uncertainties.  The \tZ, \WtZ, \ttH, three- and four-top-quark processes are
denoted $t+X$. The $WZ$, $ZZ$, $H \to ZZ$ (ggF and VBF), $HW$ and $HZ$ and VBS
processes are denoted `Bosons'.
\vspace{1ex}}

\resizebox{\columnwidth}{!}{
\begin{tabular}{%
c|
r@{\,}@{$\pm$}@{\,}l
r@{\,}@{$\pm$}@{\,}l
r@{\,}@{$\pm$}@{\,}l
r@{\,}@{$\pm$}@{\,}l
|r@{\,}@{$\pm$}@{\,}l
r@{\,}@{$\pm$}@{\,}l
|r
}
\toprule
Region &
\multicolumn{2}{c}{$t+X$} &
\multicolumn{2}{c}{Bosons} &
\multicolumn{2}{c}{Fake leptons} &
\multicolumn{2}{c|}{Total bkg.} &
\multicolumn{2}{c}{\ttW} &
\multicolumn{2}{c|}{\ttZ} &
Data \\
\midrule
\TLCR&   $0.52$ & $0.13$&    $26.9$ & $2.2$&   $2.2$ & $1.8$&    $29.5$ & $2.8$& $0.015$ & $0.004$&   $0.80$ & $0.13$&    33\\
\FLCR&   \multicolumn{2}{c}{$<0.001$} &    $39.5$ & $2.6$&   $1.8$ & $0.6$& $41.2$ & $2.7$&    \multicolumn{2}{c}{$<0.001$} &    $0.026$ & $0.007$&    39\\
\midrule
\SSLSR&    $0.94$ & $0.08$&    $0.12$ & $0.05$&    $1.5$ & $1.3$&    $2.5$ & $1.3$&   $2.32$ & $0.33$&    $0.70$ & $0.10$&    9\\
\TLSRC&    $1.08$ & $0.25$&    $0.5$ & $0.4$&    \multicolumn{2}{c}{$<0.001$} & $1.6$ & $0.5$&   $0.065$ & $0.013$&    $5.5$ & $0.7$&    8\\
\TLSRA&    $1.14$ & $0.24$&    $3.3$ & $2.2$&    $2.2$ & $1.7$&    $6.7$ & $2.8$&   $0.036$ & $0.011$&    $4.3$ & $0.6$&    7\\
\TLSRB&    $0.58$ & $0.19$&    $0.22$ & $0.18$&    \multicolumn{2}{c}{$<0.001$} &    $0.80$ & $0.26$&    $0.083$ & $0.014$&    $1.93$ & $0.28$&    4\\
\TLSRD&    $0.95$ & $0.11$&    $0.14$ & $0.12$&    $3.6$ & $2.2$&    $4.7$ & $2.2$&   $1.59$ & $0.28$&    $1.45$ & $0.20$&    10\\
\FLSRD&    $0.212$ & $0.032$&    $0.09$ & $0.07$&    $0.113$ & $0.022$& $0.42$ & $0.08$&   \multicolumn{2}{c}{$<0.001$} &    $0.66$ & $0.09$&    1\\
\FLSRE&    $0.121$ & $0.021$&    $0.07$ & $0.06$&    $0.062$ & $0.012$& $0.25$ & $0.07$&   \multicolumn{2}{c}{$<0.001$} &    $0.63$ & $0.09$&    1\\
\FLSRB&    $0.25$ & $0.04$&    $0.0131$ & $0.0032$&    $0.114$ & $0.019$& $0.37$ & $0.04$&   \multicolumn{2}{c}{$<0.001$} &    $0.75$ & $0.10$&    2\\
\FLSRC&    $0.16$ & $0.05$&    \multicolumn{2}{c}{$<0.001$} &    $0.063$ & $0.013$&   $0.23$ & $0.05$&    \multicolumn{2}{c}{$<0.001$} &    $0.64$ & $0.09$&    1\\
\bottomrule
\end{tabular}
}
\end{table}

\section{Systematic uncertainties}
\label{s:systematics}

The normalisation of signal and background in each channel can be affected by
several sources of systematic uncertainty. These are described in the
following subsections.

\subsection{Luminosity}
\label{sec:syst_lumi}
The uncertainty in the integrated luminosity in the 2015 dataset is 2.1\%. It is
derived, following a methodology similar to that detailed in
Ref.~\cite{DAPR-2011-01}, from a calibration of the luminosity
scale using $x$--$y$ beam-separation scans performed in August 2015. This
systematic uncertainty is applied to all processes modelled using \MC\
simulations.

\subsection{Uncertainties associated with reconstructed objects}
\label{sec:syst_objects}
 
Uncertainties associated with the lepton selection arise from imperfect
knowledge of the trigger, reconstruction, identification and isolation
efficiencies, and lepton momentum scale and resolution \cite{PERF-2013-03,
ATL-PHYS-PUB-2011-006, ATLAS-CONF-2014-032, ATL-PHYS-PUB-2015-041,
PERF-2015-10}.  The uncertainty in the electron identification
efficiency is the largest systematic uncertainty in the \TLC\ and among the
most important ones in the \FLC.    

Uncertainties associated with the jet selection arise from the jet energy scale
(JES), the JVT requirement and the jet energy resolution (JER).
Their estimations are based on Run-1 data and checked with early
Run-2 data.  The JES and its uncertainty are derived by combining information from
test-beam data, collision data and simulation~\cite{PERF-2012-01}.  JES
uncertainty components arising from the in-situ calibration and the jet flavour
composition are among the dominant uncertainties in the \SSLSR\ and \TL\
channels.  The uncertainties in the JER and JVT have a significant effect at
low jet \pt.  The JER uncertainty results in the second largest uncertainty in the
\TLC.

The efficiency of the flavour-tagging algorithm is measured for each jet
flavour using control samples in data and in simulation. From these
measurements, correction factors are defined to correct the tagging rates in
the simulation.  In the case of $b$-jets, correction factors and their
uncertainties are estimated based on observed and simulated $b$-tagging rates
in \ttbar dilepton events~\cite{ATLAS-CONF-2014-004}.  In the case of $c$-jets,
they are derived based on jets with identified $D^{*}$
mesons~\cite{ATLAS-CONF-2014-046}.  In the case of light-flavour jets,
correction factors are derived using dijet events~\cite{ATLAS-CONF-2014-046}.
Sources of uncertainty affecting the $b$- and $c$-tagging efficiencies are
considered as a function of jet \pt, including bin-to-bin
correlations~\cite{ATLAS-CONF-2014-004}.  An additional uncertainty is assigned
to account for the extrapolation of the $b$-tagging efficiency measurement from
the \pT region used to determine the scale factors to regions with higher \pT.
For the efficiency to tag light-flavour jets, the dependence of the uncertainty on the jet \pt and $\eta$ is considered.  These systematic uncertainties are taken as
uncorrelated between $b$-jets, $c$-jets, and light-flavour jets. 

The treatment of the uncertainties associated with reconstructed objects is common to all
three channels, and thus these are considered as correlated among different
regions.

\subsection{Uncertainties in signal modelling} 
\label{sec:signal_modeling}

From the nominal \MGAMC+\PYTHIA~8 (\tune{A14} tune) configuration, two
parameters are varied to investigate uncertainties from the modelling of the
\ttZ and \ttW processes: the renormalisation ($\mu_{\rm R}$) and factorisation ($\mu_{\rm F}$) scales.
A simultaneous variation of $\mu_{\rm R} = \mu_{\rm F}$ by factors $2.0$ and $0.5$ is performed.  In addition, the effects of a set of variations in the tune parameters
(\tune{A14} eigentune variations), sensitive to initial- and final-state
radiation, multiple parton interactions and colour reconnection, are evaluated.
Studies performed at particle level show that the largest impact comes from
variations in initial-state radiation~\cite{ATL-PHYS-PUB-2016-005}.  The
systematic uncertainty due to the choice of generator for the \ttZ and \ttW signals
is estimated by comparing the nominal sample with one generated with \SHERPA
v2.2. The \SHERPA sample uses the LO matrix element with up to one (two) additional parton(s) included
in the matrix element calculation for \ttZ  (\ttW) and merged with the \SHERPA
parton shower~\cite{Schumann:2007mg}  using the \textsc{ME+PS@LO} prescription.
The \pdf{NNPDF3\!.\!0NLO} PDF set is used in conjunction
with a dedicated parton shower tune developed by the \SHERPA authors.  Signal
modelling uncertainties are treated as correlated among channels.

\subsection{Uncertainties in background modelling}
\label{sec:bkg_modeling}

In the \TL\ and \SSLSR\ channels, the diboson background is dominated by $WZ$
production, while $ZZ$ production is dominant in the \FLC.  While the inclusive
cross sections for these processes are known to better than 10\%, they
contribute to the background in these channels if additional $b$-jets and 
other jets are produced and thus have a significantly larger uncertainty.

In the \TL\ and \SSLSR\ channels, the normalisation of the $WZ$ background is
treated as a free parameter in the fit used to extract the \ttZ and \ttW signals. The
uncertainty in the extrapolation of the $WZ$ background estimate from the
control region to signal regions with specific jet and $b$-tag multiplicities
is evaluated by comparing predictions obtained by varying the renormalisation,
factorisation and resummation scales used in MC generation. The
uncertainties vary across the different regions and an overall uncertainty of
$-50\%$ and $+100\%$ is used.

The normalisation of the $ZZ$ background is treated as a free parameter in the
fit used to extract the \ttZ and \ttW signals.  In the \FLC, several uncertainties in
the $ZZ$ background estimate are considered.  They arise from the extrapolation
from the \FLCR\ control region (corresponding to on-shell $ZZ$ production) to
the signal region (with off-shell $ZZ$ background) and from the extrapolation
from the control region without jets to the signal region with at least one
jet. They are found to be 30\% and 20\%, respectively.  An additional
uncertainty of 10--30\% is assigned to the normalisation of the heavy-flavour
content of the $ZZ$ background, based on a data-to-simulation comparison of
events with one $Z$ boson and additional jets and cross-checked with a
comparison between different $ZZ$ simulations~\cite{TOPQ-2013-05}.

The uncertainty in the \ttH background is evaluated by varying the factorisation and renormalisation scales up and down by a factor of two with respect to the nominal value, $H_{\rm T}/2$, where $H_{\rm T}$ is defined as the scalar sum of the transverse masses $\sqrt{\pt^2+m^2}$ of all final state particles.

For the \tZ background, an overall normalisation uncertainty of 50\% is
assumed.  An additional uncertainty affecting the distribution of this
background as a function of jet and $b$-jet multiplicity is evaluated by
varying the factorisation and renormalisation scales, as well as the amount of
radiation in the \tune{Perugia2012} parton shower tune.

An uncertainty of $+10\%$ and $-22\%$ is assigned to the \WtZ background cross
section. The uncertainty is asymmetric due to an alternative estimate of the
interference effect between this process and the \ttZ production.  The shape
uncertainty is evaluated by varying the factorisation and renormalisation
scales up and down by a factor of two with respect to the nominal value $H_{\rm T}/2.$

For other prompt-lepton backgrounds, uncertainties of 20\% are assigned to the
normalisations of the $WH$ and $ZH$ processes, based on calculations from
Ref.~\cite{Heinemeyer:2013tqa}.  An uncertainty of 50\% is considered for
triboson and same-sign $WW$ processes.

The fake-lepton background uncertainty is evaluated as follows.  The uncertainty due to the matrix method is estimated by propagating the
statistical uncertainty on the measurement of the fake-lepton efficiencies.
Additionally, a 20\% uncertainty is added to the subtracted charge-flip yields
estimated as the difference between data-driven charge-flips and simulation,
and the \met requirement used to enhance the single-fake-lepton fraction is varied by 
$20\,\GeV$. The main sources 
of fake muons are
decays of light-flavour or heavy-flavour hadrons inside jets.  For the \SSLSR\
region, the flavour composition of the jets faking leptons is assumed to be
unknown. To cover this uncertainty, the central values of the fake-lepton
efficiencies extracted from the $b$-veto and the $b$-tag control regions are
used, with the efficiency difference assigned as an extra uncertainty. For the
\FL\ channel, fake-lepton systematic uncertainties are covered by the
scale-factor uncertainties used to calibrate the simulated fake-lepton yield in
the control regions. Within a fake-lepton estimation method, all systematic
uncertainties are considered to be correlated among analysis channels and
regions. Thus \SSLSR\ and \TL\ fake-lepton systematic uncertainties that use the
matrix method are not correlated with the \FL\ systematic uncertainties.  The
expected uncertainties in the fake-lepton backgrounds relative to the total
backgrounds vary in each channel and signal region: 50\% for the \SSLSR\
region, 25--50\% for the \TLC\ and 5--10\% for the \FLC.

\section{Results}
\label{s:results}

In order to extract the \ttZ and \ttW cross sections, nine signal regions (\SSLSR, 
\TLSRA, \TLSRB, \TLSRC, \TLSRD, \FLSRB, \FLSRC, \FLSRD, \FLSRE) and two control regions
(\TLCR, \FLCR) are simultaneously fitted. 
The \SSLSR\ signal region is particularly sensitive to \ttW, the \TLSRD\ signal region 
is sensitive to both, \ttW and \ttZ, while all other signal regions aim at the determination
of the \ttZ cross section.
The cross sections \sttZ and \sttW are determined using a binned maximum-likelihood fit
to the numbers of events in these regions. 
The fit is based on the profile-likelihood
technique, where systematic uncertainties are allowed to vary as nuisance
parameters and take on their best-fit values.  None of the uncertainties are
found to be significantly constrained or pulled from their initial values.  The calculation of confidence intervals and
hypothesis testing is performed using a modified frequentist method as implemented in RooStats~\cite{RooFit,RooFitManual}.

\begin{figure}[htbp]
\centering
\includegraphics[width=0.85\textwidth]{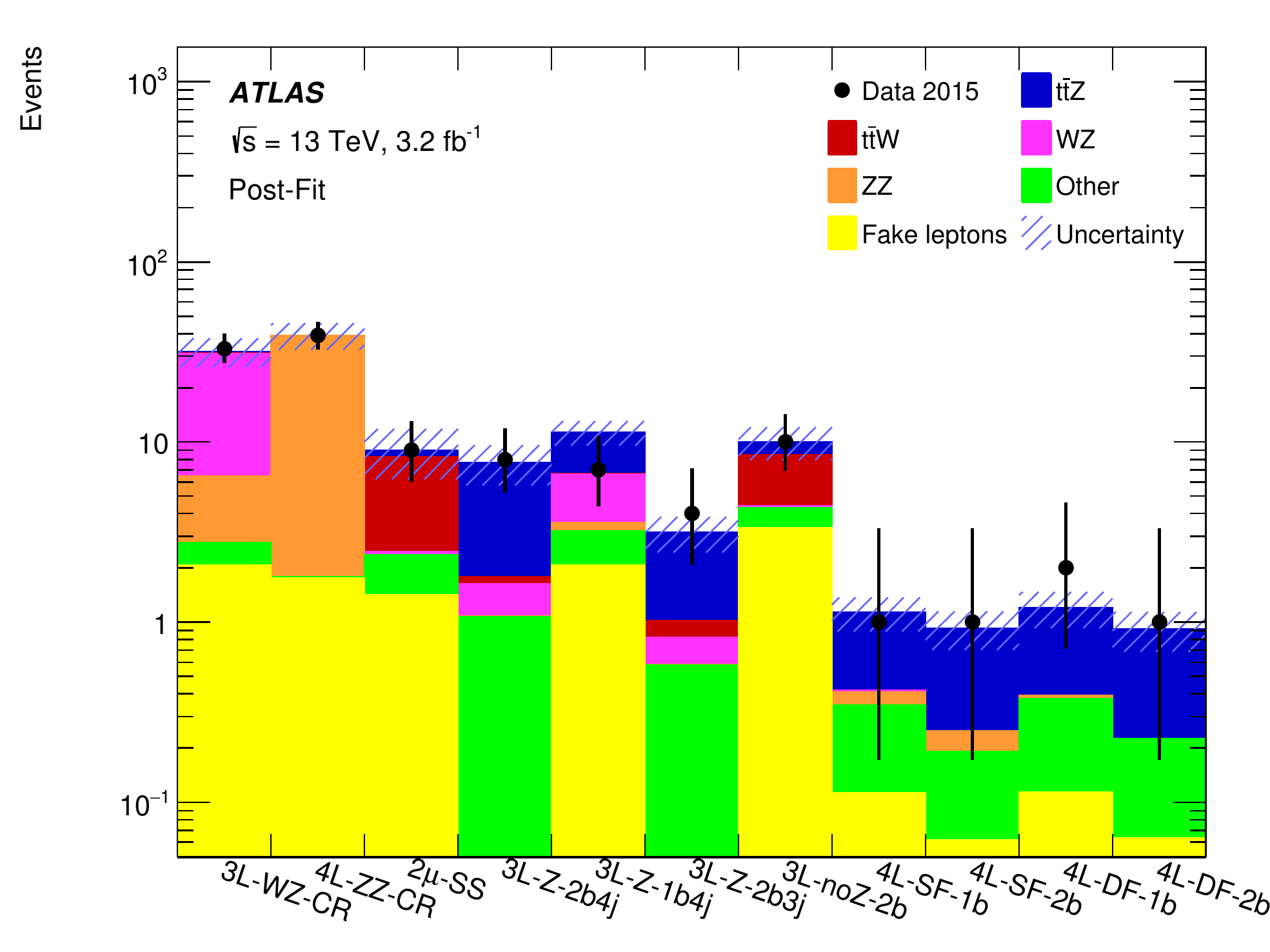}
\caption{Expected yields after the fit compared to data for the
fit to extract \sttZ and \sttW in the signal regions and in the control regions
used to constrain the $WZ$ and $ZZ$ backgrounds.  The `Other' background
summarises all other backgrounds described in Section~\ref{s:samples}.  
\hatch.\label{fig:allyields}}
\end{figure}

A summary of the fit to all regions used to measure the \ttZ and \ttW
production cross sections are shown in Figure~\ref{fig:allyields}.  The
normalisation corrections for the $WZ$ and $ZZ$ backgrounds with respect to the
Standard Model predictions are obtained from the fits as described in
Section~\ref{s:selection} and found to be compatible with unity: 
$1.11 \pm 0.30$
for the $WZ$ background and
$0.94 \pm 0.17$
for the $ZZ$ background.
\vspace{1ex}

\begin{figure}[htbp]
\centering
\includegraphics[width=0.85\textwidth]{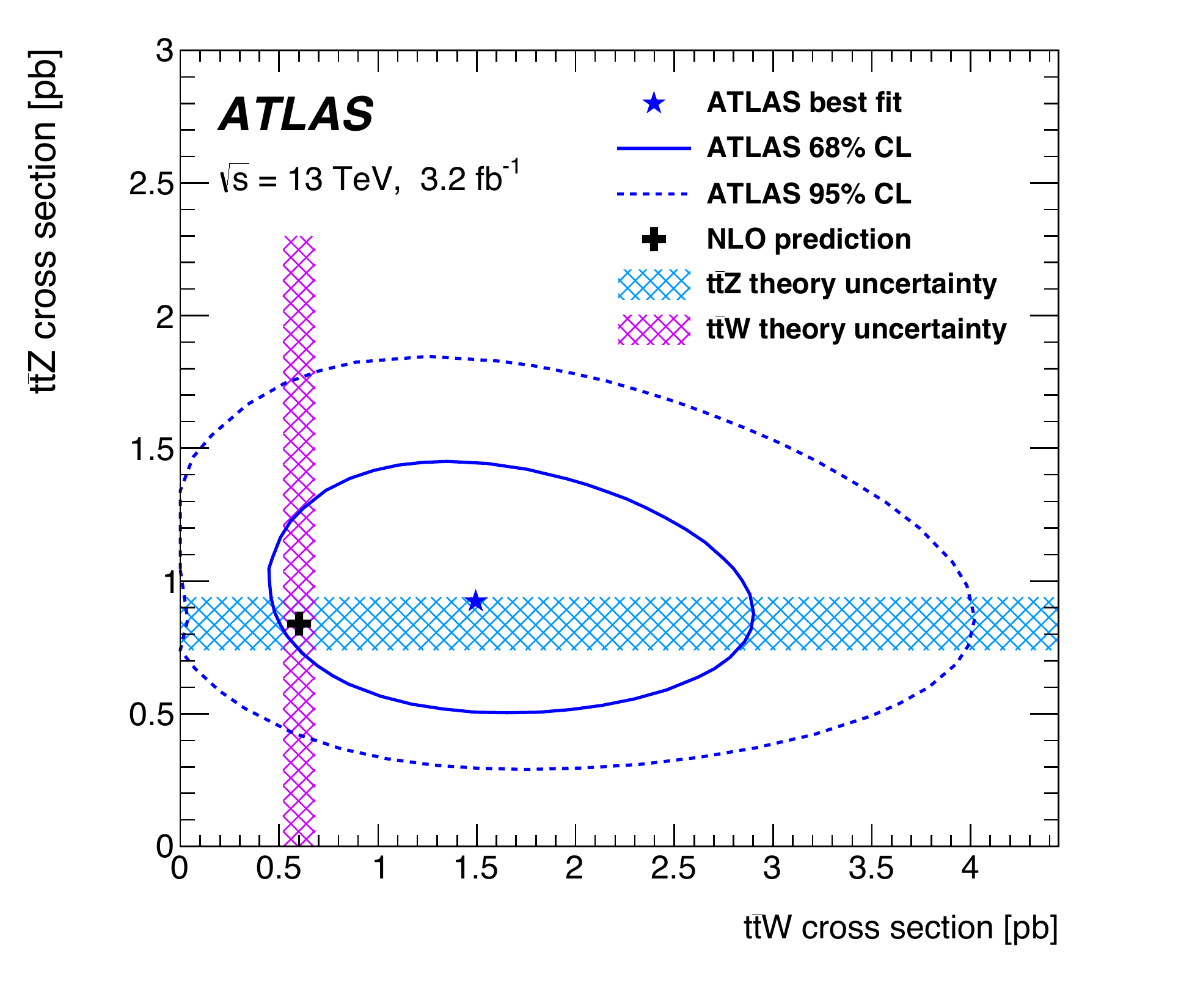}
\caption{The result of the simultaneous fit to the \ttZ and \ttW cross sections
along with the 68\% and 95\% confidence level (CL) contours. The shaded areas
correspond to the theoretical uncertainties in the Standard Model predictions,
and include renormalisation and factorisation scale uncertainties as well as
PDF uncertainties including $\alphas$ variations.}
\label{fig:simfit}
\end{figure}

The results of the fit are $\sigma_{\ttZ} = 0.92 \pm 0.29 \stat \pm 0.10 \syst$
pb and $\sigma_{\ttW} = 1.50 \pm 0.72 \stat \pm 0.33 \syst$ pb with a
correlation of $-0.13$ and are shown in Figure~\ref{fig:simfit}.  The fit
yields significances of $3.9\sigma$ and $2.2\sigma$ over the background-only
hypothesis for the \ttZ and \ttW processes, respectively. The expected
significances are $3.4\sigma$ for \ttZ and $1.0\sigma$ for \ttW production.
The significance values are computed using the asymptotic approximation described in Ref.~\cite{cls_3}. 
In the two channels most sensitive to the \ttW signal the observed relative
number of events with two positively or two negatively charged leptons
is compatible with expectation. In the \TLSRD\ channel the observed distribution of 
the number of events with a given amount of electrons and muons match expectation, as well.

Table~\ref{tab:syst} shows the leading and total uncertainties in the
measured \ttZ and \ttW cross sections. In estimating the uncertainties for \ttZ (\ttW), the cross section for \ttW (\ttZ) is fixed to its
Standard Model value. For both processes, the precision of
the measurement is dominated by statistical uncertainties.  For the \ttZ determination,
the different sources contribute with similar size to the total systematic uncertainty.
For the \ttW determination, the dominant systematic uncertainty source is
the limited amount of data available for the estimation of the fake leptons.

\begin{table}[htbp]
\centering \renewcommand{\arraystretch}{1.2}
\caption{List of dominant and total uncertainties in the measured cross 
sections of the \ttZ and \ttW processes from the fit. All uncertainties are
symmetrised.\vspace{1ex}}
\label{tab:syst}
\begin{tabular}{lcc}
\toprule
Uncertainty                 &   $\sigma_{\ttZ}$ & $\sigma_{\ttW}$ \\
\midrule
Luminosity                    & 2.6\% & 3.1\%  \\
Reconstructed objects         & 8.3\% & 9.3\%  \\
Backgrounds from simulation   & 5.3\% & 3.1\%  \\
Fake leptons and charge misID & 3.0\% & 19\%  \\
Signal modelling   & 2.3\% & 4.2\%  \\
\midrule
Total systematic            & 11\% & 22\% \\
Statistical                 & 31\% & 48\% \\
\midrule
Total                       & 32\% & 53\% \\
\bottomrule
\end{tabular}
\end{table}

\newpage
\section{Conclusion}
\label{s:conclusion}

Measurements of the production cross sections of a top-quark pair in
association with a $Z$ or $W$ boson using \lumi of data collected by the ATLAS
detector in $\sqrt{s} = 13\,\TeV$ $pp$ collisions at the LHC are presented. 
Final states with either two same-charge muons, or three or four leptons
are analysed.  From a simultaneous fit to nine signal regions and two control regions, the
\ttZ and \ttW production cross sections are determined to be $\sttZ = 0.9 \pm
0.3$ pb and $\sttW = 1.5 \pm 0.8$ pb. 
Both measurements are consistent with the NLO
QCD theoretical calculations, $\sttZ = 0.84 \pm 0.09\,\text{pb}$ and $\sttW =
0.60 \pm 0.08\,\text{pb}$.

\section*{Acknowledgements}

We thank CERN for the very successful operation of the LHC, as well as the
support staff from our institutions without whom ATLAS could not be
operated efficiently.

We acknowledge the support of ANPCyT, Argentina; YerPhI, Armenia; ARC, Australia; BMWFW and FWF, Austria; ANAS, Azerbaijan; SSTC, Belarus; CNPq and FAPESP, Brazil; NSERC, NRC and CFI, Canada; CERN; CONICYT, Chile; CAS, MOST and NSFC, China; COLCIENCIAS, Colombia; MSMT CR, MPO CR and VSC CR, Czech Republic; DNRF and DNSRC, Denmark; IN2P3-CNRS, CEA-DSM/IRFU, France; GNSF, Georgia; BMBF, HGF, and MPG, Germany; GSRT, Greece; RGC, Hong Kong SAR, China; ISF, I-CORE and Benoziyo Center, Israel; INFN, Italy; MEXT and JSPS, Japan; CNRST, Morocco; FOM and NWO, Netherlands; RCN, Norway; MNiSW and NCN, Poland; FCT, Portugal; MNE/IFA, Romania; MES of Russia and NRC KI, Russian Federation; JINR; MESTD, Serbia; MSSR, Slovakia; ARRS and MIZ\v{S}, Slovenia; DST/NRF, South Africa; MINECO, Spain; SRC and Wallenberg Foundation, Sweden; SERI, SNSF and Cantons of Bern and Geneva, Switzerland; MOST, Taiwan; TAEK, Turkey; STFC, United Kingdom; DOE and NSF, United States of America. In addition, individual groups and members have received support from BCKDF, the Canada Council, CANARIE, CRC, Compute Canada, FQRNT, and the Ontario Innovation Trust, Canada; EPLANET, ERC, FP7, Horizon 2020 and Marie Sk{\l}odowska-Curie Actions, European Union; Investissements d'Avenir Labex and Idex, ANR, R{\'e}gion Auvergne and Fondation Partager le Savoir, France; DFG and AvH Foundation, Germany; Herakleitos, Thales and Aristeia programmes co-financed by EU-ESF and the Greek NSRF; BSF, GIF and Minerva, Israel; BRF, Norway; Generalitat de Catalunya, Generalitat Valenciana, Spain; the Royal Society and Leverhulme Trust, United Kingdom.

The crucial computing support from all WLCG partners is acknowledged gratefully, in particular from CERN, the ATLAS Tier-1 facilities at TRIUMF (Canada), NDGF (Denmark, Norway, Sweden), CC-IN2P3 (France), KIT/GridKA (Germany), INFN-CNAF (Italy), NL-T1 (Netherlands), PIC (Spain), ASGC (Taiwan), RAL (UK) and BNL (USA), the Tier-2 facilities worldwide and large non-WLCG resource providers. Major contributors of computing resources are listed in Ref.~\cite{ATL-GEN-PUB-2016-002}.

\printbibliography
\newpage
\begin{flushleft}
{\Large The ATLAS Collaboration}

\bigskip

M.~Aaboud$^\textrm{\scriptsize 137d}$,
G.~Aad$^\textrm{\scriptsize 88}$,
B.~Abbott$^\textrm{\scriptsize 115}$,
J.~Abdallah$^\textrm{\scriptsize 66}$,
O.~Abdinov$^\textrm{\scriptsize 12}$,
B.~Abeloos$^\textrm{\scriptsize 119}$,
R.~Aben$^\textrm{\scriptsize 109}$,
O.S.~AbouZeid$^\textrm{\scriptsize 139}$,
N.L.~Abraham$^\textrm{\scriptsize 153}$,
H.~Abramowicz$^\textrm{\scriptsize 157}$
% ...
\bigskip
\\
% ...
$^{12}$ Institute of Physics, Azerbaijan Academy of Sciences, Baku, Azerbaijan\\
% ...
$^{66}$ University of Iowa, Iowa City IA, United States of America\\
% ...
$^{88}$ CPPM, Aix-Marseille Universit{\'e} and CNRS/IN2P3, Marseille, France\\
% ...
$^{109}$ Nikhef National Institute for Subatomic Physics and University of Amsterdam, Amsterdam, Netherlands\\
% ...
$^{115}$ Homer L. Dodge Department of Physics and Astronomy, University of Oklahoma, Norman OK, United States of America\\
% ...
$^{119}$ LAL, Univ. Paris-Sud, CNRS/IN2P3, Universit{\'e} Paris-Saclay, Orsay, France\\
% ...
$^{137}$ $^{(a)}$ Facult{\'e} des Sciences Ain Chock, R{\'e}seau Universitaire de Physique des Hautes Energies - Universit{\'e} Hassan II, Casablanca; $^{(b)}$ Centre National de l'Energie des Sciences Techniques Nucleaires, Rabat; $^{(c)}$ Facult{\'e} des Sciences Semlalia, Universit{\'e} Cadi Ayyad, LPHEA-Marrakech; $^{(d)}$ Facult{\'e} des Sciences, Universit{\'e} Mohamed Premier and LPTPM, Oujda; $^{(e)}$ Facult{\'e} des sciences, Universit{\'e} Mohammed V, Rabat, Morocco\\
% ...
$^{139}$ Santa Cruz Institute for Particle Physics, University of California Santa Cruz, Santa Cruz CA, United States of America\\
% ...
$^{153}$ Department of Physics and Astronomy, University of Sussex, Brighton, United Kingdom\\
% ...
$^{157}$ Raymond and Beverly Sackler School of Physics and Astronomy, Tel Aviv University, Tel Aviv, Israel\\
% ...
$^{a}$ Also at Department of Physics, King's College London, London, United Kingdom\\
$^{b}$ Also at Institute of Physics, Azerbaijan Academy of Sciences, Baku, Azerbaijan\\
$^{c}$ Also at Novosibirsk State University, Novosibirsk, Russia\\
$^{d}$ Also at TRIUMF, Vancouver BC, Canada\\
$^{e}$ Also at Department of Physics {\&} Astronomy, University of Louisville, Louisville, KY, United States of America\\
% ...
\end{flushleft}



\end{document}